\documentclass{IEEEtran}
\usepackage[utf8]{inputenc}

\usepackage{amsmath,amssymb,amsfonts,amsthm,tikz,caption,subcaption}
\usepackage{textcomp}
\usepackage{cite}
\usepackage{threeparttable}
\usepackage{array}
\usepackage{float}

\usepackage{tikz}
	\usetikzlibrary{shapes,decorations,decorations.pathreplacing}
	\usetikzlibrary{arrows,shapes,positioning,calc,fadings,patterns}
	\usetikzlibrary{decorations,decorations.pathreplacing,decorations.pathmorphing}
\usepackage{tikz-3dplot}

\newcolumntype{C}[1]{>{\centering\let\newline\\\arraybackslash\hspace{0pt}}m{#1}}

\usepackage{soul,xcolor}

\title{An Overview of Machine Learning Techniques for Radiowave Propagation Modeling}
\author{Aristeidis Seretis, Costas D. Sarris}
\date{December 2019}

\begin{document}

\setstcolor{red}

\maketitle

\begin{abstract}

We give an overview of recent developments in the modeling of radiowave propagation, based on machine learning algorithms.  We identify the input and output specification and the architecture of the model as the main challenges associated with machine learning-driven propagation models. Relevant papers are discussed and categorized based on their approach to each of these challenges. Emphasis is given on presenting the prospects and open problems in this promising and rapidly evolving area.  

\end{abstract}

\begin{IEEEkeywords}

Artificial Intelligence, Machine learning, Neural Networks, Radiowave Propagation, Propagation Losses

\end{IEEEkeywords}

\section{Introduction}\label{sec_1}

For the intelligent planning and efficient management of any wireless communication system, channel propagation models are indispensable \cite{catedra}. As a growing number of wireless services with high performance demands is offered, the need for new propagation models becomes more urgent. Safety-critical, high-throughput and low-latency are just some of the required characteristics needed in current and future wireless systems.

Over the years, various empirical propagation models, such as Okumura-Hata or Walfish-Bertoni among others, have been created \cite{molisch, rappaport}. Empirical models are measurement-driven, formulated after fitting measurements taken at a specific site. These models are computationally efficient, yet they fail to capture the full spectrum of complex wave effects that often determine the performance of a wireless link.

On the other hand, deterministic models 
are increasingly used over the past years. Such models include methods based on solving Maxwell's equations, such as integral equation \cite{fem} and finite difference \cite{fdtd} methods. Moreover, approximate methods, such as ray-tracing (RT) for indoor and urban scenarios \cite{rt}, and the vector parabolic equation (VPE) method for terrestrial propagation and propagation in tunnels \cite{vpe}, have been popular deterministic techniques. Deterministic models are site-specific. Therefore, they can provide reliable predictions for a given environment. Nevertheless, despite recent advances in processing power, their computational demands are still considered high.

Machine learning (ML)-driven propagation models are promising tools for resolving the standard dichotomy between accuracy and efficiency of propagation models. They can be trained offline, by either measured or synthetic (simulated) data. Moreover, their highly non-linear nature makes them promising candidates for predicting propagation parameters such as multipath fading. Finally, they can be made either site-specific or general-purpose, something that gives them great flexibility.

Given the growing interest in ML techniques for propagation modeling, we present an overview of various relevant papers. We discuss what an ML-driven propagation model is and how it can be created. We also focus on explaining how a propagation scenario can influence various decisions regarding the ML propagation model. 

The paper is organized as follows. First, in Section II, the three main building blocks of any ML radio propagation model are introduced. These are the input to the ML model, the model itself, and its output. Then, the challenges associated with each one of them, as dealt with by various ML-based propagation modeling papers, are discussed. The key ideas drawn from these papers are presented in the next three sections. Section III identifies several ways to specify the input to the ML model. Section IV highlights key points regarding the various ML models that have been used for propagation modeling, while Section V presents the types of output data that have been derived through these models. Section VI presents the main conclusions of the paper.

\section{ML propagation models}\label{sec_2}

\subsection{ML propagation models: goal}

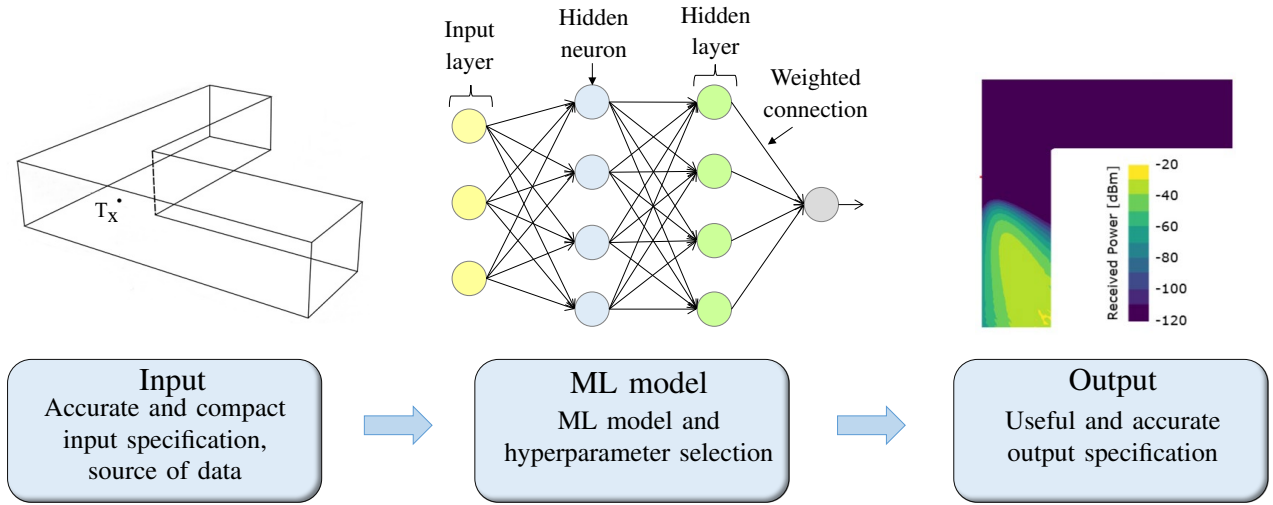
\begin{figure*}\centering
\vspace*{-0.4cm}
\begin{tikzpicture}
\node at(0,0){\includegraphics[scale=0.5,clip,trim={0cm 0cm 0cm 0cm}]{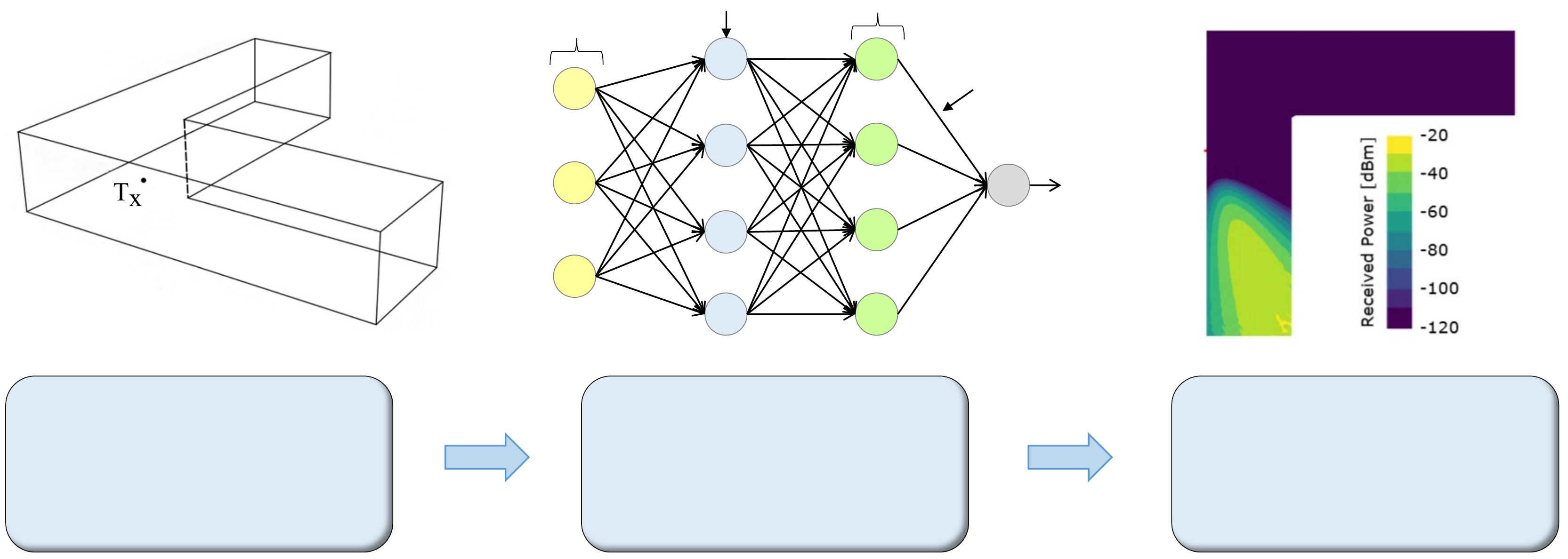}};

\node at(-6.2, -1.35){\large{Input}};
\node at(-0, -1.35){\large{ML model}};
\node at(6.3, -1.35){\large{Output}};

\node at(-0.6, 3.3)[align=center]{\small{Hidden} \\ \small{neuron}};;

\node at(-2.25, 3.1)
[align=center]{\small{Input} \\ \small{layer}};

\node at(1.02, 3.3)
[align=center]{\small{Hidden} \\ \small{layer}};

\node at(2.35, 2.5)
[align=center]{\small{Weighted} \\ \small{connection}};

\node at(-6.3, -2.1)[align=center]{ Accurate and compact \\ input specification, \\ source of data};

\node at(-0, -2.1)[align=center]{ ML model and \\ hyperparameter selection };

\node at(6.3, -2.1)[align=center]{ Useful and accurate \\ output specification };

\end{tikzpicture}
\caption{Flowchart of an ML-driven propagation model, along with its main challenges.}
\label{fig:propagation_flowchart}
\vspace*{-0.4cm}
\end{figure*}

Generally, ML problems can be classified into supervised and unsupervised. 
Supervised problems are associated with data pairs of ($\boldsymbol{x}$, $\boldsymbol{y}$), where the input $\boldsymbol{x}$ to the ML model is mapped to a specific output $\boldsymbol{y}$. On the other hand, in unsupervised problems, only the input $\boldsymbol{x}$ is known. For supervised problems, the main goal of the ML model is to learn an unknown function $f$, mapping an input space {$V_x$} to a target space {$V_y$}:
\begin{equation}
    f:V_x\xrightarrow{}V_y
   \label{eq:1} 
\end{equation}
where $V_x$ is the set of all possible input vectors $\boldsymbol{x}$ and $V_y$ is the set of all possible output vectors $\boldsymbol{y}$, accordingly. The ML model, however, does not have access to the whole set of $V_x$ and $V_y$, but attempts to approximate $f$ with a function $g$ that is computed from the given data pairs of $(\boldsymbol{x},\boldsymbol{y})$:
\begin{equation}
    g:\boldsymbol{x}\xrightarrow{}\boldsymbol{y}
   \label{eq:2} 
\end{equation}
The iterative process of computing $g$ is called training. It revolves around minimizing the in-sample error between the ML model's predictions $g(\boldsymbol{x})$ and the true target values $\boldsymbol{y}$, given a cost function $L$, over the parameters of the ML model $\boldsymbol{\psi}$: 
\begin{equation}
\underset{\boldsymbol{\psi}}{\text{min}} \, L(\mathcal{E}_{\text{in}})
\end{equation} 
At the end of the training procedure, the model has learned parameters $\hat{\boldsymbol{\psi}}$, so that the final output of the model is:
\begin{equation}
    \hat{\boldsymbol{y}} = g(\boldsymbol{x}|\hat{\boldsymbol{\psi}})
   \label{eq:3} 
\end{equation} 
For the unsupervised problems, the purpose of the ML model is to find underlying patterns in the input data, e.g. grouping them into classes. Given these definitions, computing various propagation parameters, such as the path loss (PL) or the received signal strength (RSS) at a location, from measured or synthetic data, is a supervised problem.


One of the most important attributes of any ML model is its ability to generalize to similar problems, rather than using its parameters to memorize a specific one. 
An ML model is expected to not only exhibit low in-sample error, but small out-of sample error as well, computed on data not used during its training. In the context of PL prediction, an ML propagation model should be sufficiently accurate when tested on data collected at different environments or taken at different positions than the ones used to train it. Since $f$ is unknown, $g$ is only an approximation of $f$ computed on a dataset that may also contain noisy samples. Thus, it can be proved that \cite{ml_book}:
\begin{equation}
    \mathcal{E}_{\text{out}}(g) \leq \mathcal{E}_{\text{in}}(g) + \Omega(g) 
   \label{eq:gen_bound} 
\end{equation}
where $\Omega(g)$ is a complexity penalty term associated with the final model $g$ that has been chosen as part of the training process. Eq.~(\ref{eq:gen_bound}) implies that for the ML model to have good generalization abilities, the in-sample error has to be made as small as possible, while also keeping the model complexity to a minimum. Simple models may not be capable of achieving a small enough $\mathcal{E}_{\text{in}}$, underfitting the available data. Overly complex models can potentially achieve a zero $\mathcal{E}_{\text{in}}$ at the cost of a large complexity penalty term, overfitting the data. This is known as the bias-variance trade-off \cite{ml_book}. Finding a balance between minimizing $\mathcal{E}_{\text{in}}$ and restricting $\Omega(g)$ is accomplished through the evaluation of different models on a dataset that is not used during training. Consequently, it is common practice to create three different datasets for a specific task; the training, the validation and the test set \cite{ml_book}. The training set is used to train the learning parameters $\boldsymbol{\psi}$ of a network, such as the weights $\boldsymbol{w}$ and biases $\boldsymbol{b}$ in in an artificial neural network (ANN). The validation set is used for model selection as well as for ensuring that the network is not overfitting during training. Finally, the test set is used to evaluate the chosen ML model, after it is trained.

\subsection{ML propagation models: challenges}

The general flowchart of an ML-based propagation model can be seen in Fig.~\ref{fig:propagation_flowchart}. By inspection of the diagram, the three building blocks of any ML propagation model, along with their main challenges, can be identified.

\subsubsection{Input}

The input to the ML model should contain features that are relevant to the propagation problem and representative of the relation between $\boldsymbol{x}$ and $\boldsymbol{y}$. It has to also be compact to avoid long training times. The input data can be measured, creating a direct connection between the trained model and the reality observed, or synthetic ones, generated by site-specific or empirical models (RT, VPE, PL exponent (PLE) models etc).

\subsubsection{ML model}

The choice of the ML model defines the type of function $g$ that we seek to learn. The function is often non-linear, since that gives the model additional degrees of freedom to fit the data. There are many available ML models, from deep ANNs \cite{goodfellow} to powerful implementations of regression decision trees and support vector machines (SVMs) \cite{hastie}. The hyperparmeters of the ML model have to also be carefully chosen \cite{hyper}. Those include parameters that strongly affect the model’s performance, although they are pre-fixed rather than 
trained.

\subsubsection{Output}

Finally, the output specification of the ML model has to contain useful information about the propagation characteristics of the communication channel. The output can be a scalar quantity $y$, such as the PLE of the communication channel, or a vector $\boldsymbol{y}$ consisting of complex electromagnetic field values, at one or more receiver points. The output can also represent a probabilistic PL model. Finally, the predictions of the ML model have to exhibit small  $\mathcal{E}_{\text{out}}$.

In the following, we group the papers under review in three categories based on their approach to the three challenges we identified (input, ML model and output). Each category is further divided in sub-categories to better reflect the diversity of the work that has been conducted in the area. We also present a brief, yet representative case study of how to create an ML propagation model in the Appendix.

\section{Input specification for ml models}

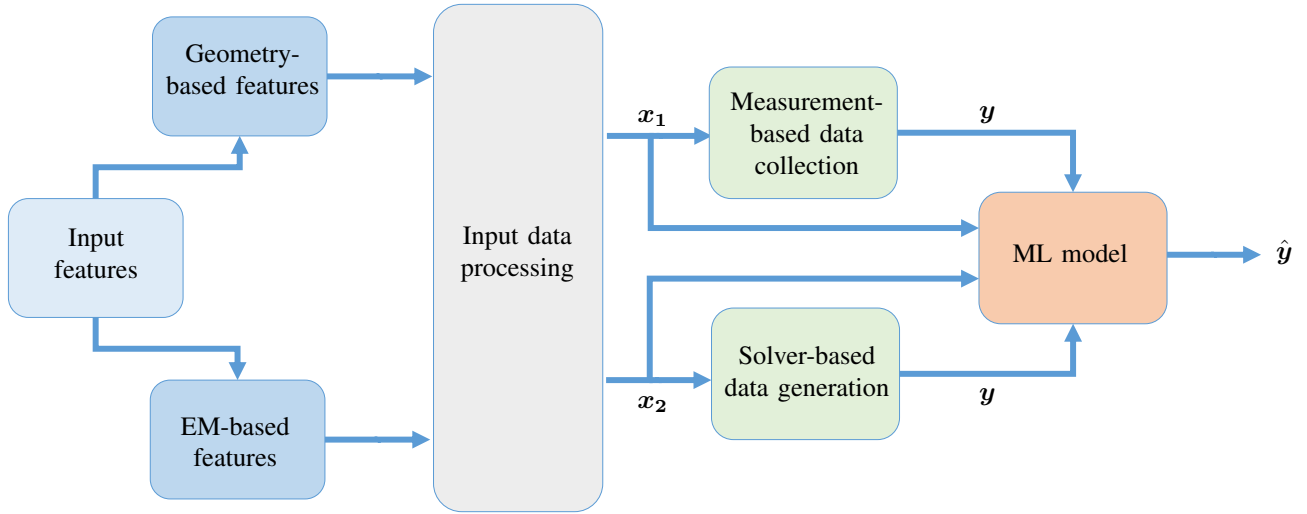
\begin{figure*}[h!]\centering
\vspace*{-0.4cm}
\begin{tikzpicture}
\node at(0,0){\includegraphics[scale=0.5,clip,trim={0cm 0cm 0cm 0cm}]{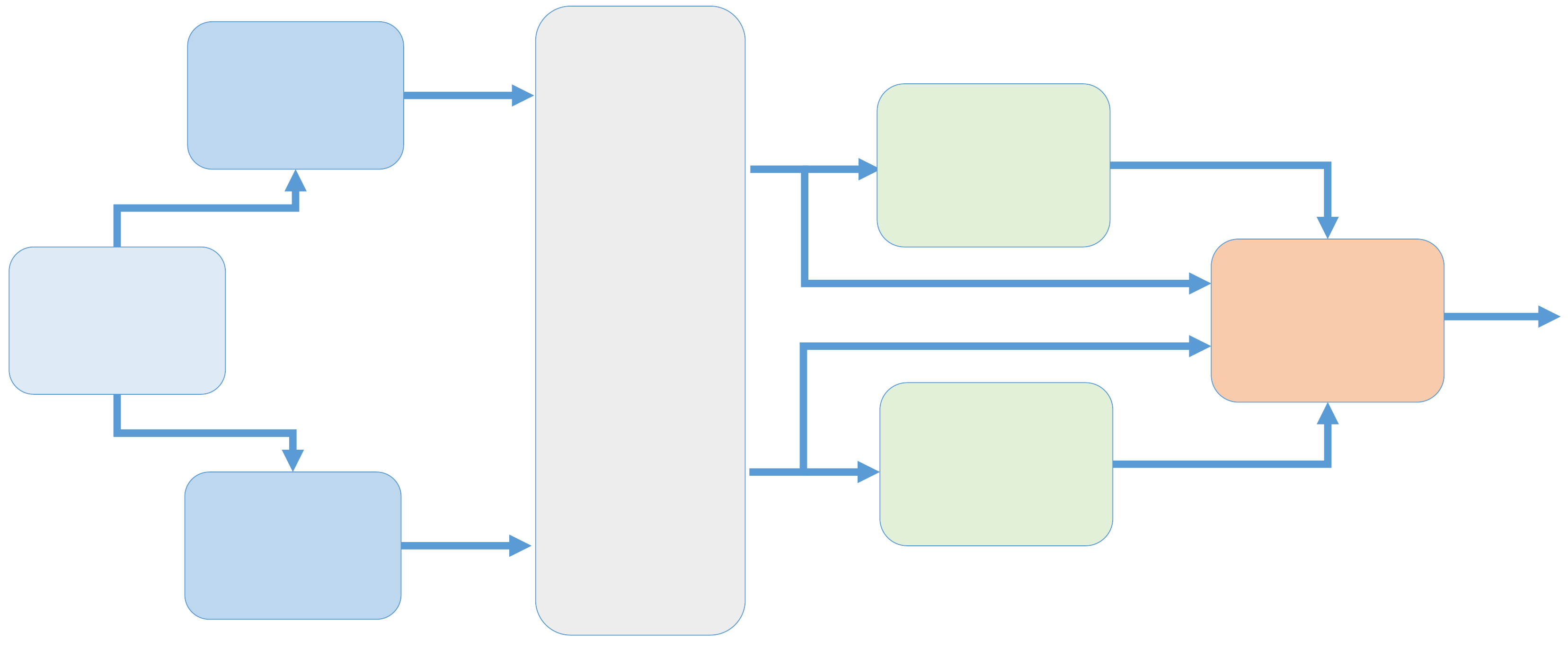}};

\node at(-5.2, 2.55) [align=center]
{Geometry- \\ based features};

\node at(-5.3, -2.4) [align=center]
{EM-based \\ features};

\node at(-7.15, 0.1) [align=center]
{Input \\ features};

\node at(-1.55, 0.1) [align=center]
{Input data \\ processing};

\node at(5.8, 0.1) {ML model};

\node at(8.65, 0.125){$\hat{\boldsymbol{y}}$};

\node at(0.25, 1.9){$\boldsymbol{x_1}$};

\node at(0.25, -1.9) {$\boldsymbol{x_2}$};

\node at(4.7, 1.95){$\boldsymbol{y}$};
\node at(4.7, -1.8){$\boldsymbol{y}$};

\node at(2.3, 1.7) [align=center]
{Measurement- \\ based data \\ collection};

\node at(2.3, -1.5) [align=center]
{Solver-based \\ data generation};

\end{tikzpicture}
\caption{The flowchart of the input specification of an ML model.}
\label{fig:input_specification}
\vspace*{-0.4cm}
\end{figure*}

The general flowchart for generating the input to the ML model can be seen in Fig.~\ref{fig:input_specification}. Input features specify the geometry (topographic information of an area or indoor floorplan, position of antennas) and electromagnetic properties of a communication channel (pattern/polarization of antennas, permittivity/conductivity of various surfaces, frequency of operation). The input features have to be pre-processed before they are used by the ML model. That processing step may include feature scaling and normalization or dimensionality reduction techniques, among others \cite{hastie}. After processing the input data, input vectors $\boldsymbol{x}$ are generated. Depending on how the target values $\boldsymbol{y}$ are created, i.e via measurements or through a model, the input vectors may be different, hence the difference in notation ($\boldsymbol{x_1}$ and $\boldsymbol{x_2}$, respectively). Either way, measured or generated $\boldsymbol{y}$ vectors will be used together with $\boldsymbol{x_1}$ or $\boldsymbol{x_2}$ to form the training pairs. After training, the ML model is able to output its prediction $\boldsymbol{\hat{y}}$ for any new input vector $\boldsymbol{x}$.

\subsection{Modeling environment}

Over the past years, many papers have used an ML approach for determining various large-scale propagation characteristics, such as path gain (PG) or PLE, for a variety of diverse communication environments. There have been papers focusing on urban environments, such as \cite{neskovic, chang}, rural, such as \cite{ostlin, cheng1}, or even a mix of different outdoor environments, such as \cite{ayadi}. Special environments such as roads, mines and subway tunnels have also been considered \cite{zaarour, di_wu, aris}.

\subsubsection{Environmental and topographical features}

The propagation environment plays an important role and can influence many design choices for the ML propagation model, such as what input features to use. As an example, parameters such as the number and width of the streets, the height of the buildings, the building separation and the orientation of the streets are often used in urban environments \cite{neskovic, popescu_2}. In forested areas, some input features may relate to the vegetation and canopy in the environment \cite{oroza}. For propagation over irregular terrain, the path profile can be used as an input to the ML model \cite{ribero}. A path profile is created by tracing the line connecting the receiver and the transmitter, sampling the elevation of the ground at fixed intervals. This is done to account for the morphological variations of the ground. It can also be used as an input in urban scenarios, where there may be numerous obstacles obstructing line of sight (LOS) \cite{piacentini}.


\subsubsection{Propagation features}

The input to the ML model can also take into account the diverse propagation mechanisms present in an environment. A common trend among papers modeling propagation in urban areas is to differentiate between LOS and non-LOS (NLOS) cases \cite{neskovic, popescu_2}. For NLOS cases, the authors often use an expanded input set to account for their higher complexity. This differentiation has been shown to improve the accuracy of the model \cite{popescu_2}. Moreover, several papers on urban propagation also include input features that account for diffraction losses \cite{fraile, balandier, ferreira}. Diffraction is more pronounced in such environments. Hence, its inclusion generally improves the accuracy of the ML model, irrespective of its type. In \cite{balandier}, the authors investigated the influence diffraction losses can have in the accuracy of their ML propagation model (ANN). They found that the ANN that accounted for diffraction losses in the city of Paris was more accurate than one that did not. Similar findings were also reported in \cite{popescu_2}.


\subsection{Input features}


\subsubsection{Type of input features}

For most propagation modeling cases, input features take continuous, real values, such as the frequency of operation or the distance between transmitter and receiver. Input features can also take discrete or even binary values. For example, the $j$-th input feature $x^{j}$ of input vector $\boldsymbol{x}$ may be binary, where $x^{j}=\{0,1\}$, denoting the presence or absence of an LOS component. Additionally, there may be an input feature for classifying the type of environment in the vicinity of the receiver \cite{neskovic}, \cite{lina_wu}, or the terrain complexity \cite{oroza}. For example, we can have $x^{j}=\{0, 1, ... , M-1\}$, where $M$ represents the number of different types/classes of the feature. The input to the ML model can also be visual, as in \cite{thrane, ahmadien}, where the authors utilized satellite images as part of their training data.

When no correlation between different samples is assumed, training samples can be used as individual inputs to the ML model. When there is dependence among different samples, they can also be passed on to the ML model as sequences of input data. The length of the sequence is a hyperparameter that has to be tuned accordingly. If it is set too small for a given problem, the network may not fully exploit the correlation between different samples. On the contrary, if the length is set too high, the different sequences may be uncorrelated. That can lead to unnecessary computations or to a sub-optimal ML model. For example, in \cite{xu}, the authors found that using a sequence of 200 RSS samples collected at different timestamps gave better results than using a larger or smaller number of them. Similar findings were reported in \cite{adege, hoang, wang}.


\subsubsection{Number of input features}

Expanding the input information given to the ML model by increasing the number of input features, when uncorrelated, generally improves the model predictions. In \cite{gschwendtner}, the authors found that their ML model predictions improved, when the number of inputs was increased. In \cite{sotiroudis_1}, the authors used an RT solver to simulate propagation in an urban environment. They provided the network with global information with regards to the height of the building at the center of each cell, as well as the transmitter and receiver coordinates. They also provided local information to the ANN, i.e. the same type of input data as the global one, but using only 8 building heights in the proximity of the receiver. The model that used global and local information was more accurate than the one that used only local information.

\subsubsection{Dimensionality reduction}

When the number of input parameters is increased disproportionately with respect to the underlying complexity of the problem, the computational performance of the network is compromised. For these cases, dimensionality reduction techniques can be helpful \cite{adege}. For example, in \cite{piacentini}, the authors discretized the path between the transmitter and the receiver. Each one of the discrete segments was assigned input variables describing the main obstacle present there. Two dimensionality reduction techniques were used to reduce the input space: principal component analysis (PCA) and a nonlinear PCA (nPCA). The authors found that reducing the input representation helped considerably. In cases where a path profile is generated, PCA is readily used to condense the high-dimensional input space into a lower-dimensional. A similar procedure was followed in a number of other papers, such as \cite{lina_wu}, where the authors used PCA to convert 9 input features describing the surrounding environment into 4 principal components, and \cite{jo}, where the authors mapped 4 input features to one. Dimensionality reduction techniques are also good candidates for reducing correlation among the input parameters. Even though some ML models, like ANNs, are highly immune to redundant information, using more inputs than needed affects the computational performance of the model. For example, \cite{lina_wu} and \cite{jo} showed that dimensionality reduction accounted for up to 30\% savings in training time, for comparable model accuracy. Reducing input dimensionality can also be achieved without explicitly using PCA techniques. In \cite{thrane}, the authors converted the colored input images of a propagation area into grey-scale, thus decreasing the number of color channels from 3 to 1.

\subsubsection{Impact of input features}

It is often necessary to choose the input parameters by trying different options. In \cite{popescu_1}, the authors tried different numbers of input features. They saw bigger improvements in the accuracy of their radial basis function (RBF)-ANN, by adding street orientation as an input feature, compared to adding the difference in height between the base station and the buildings, for urban and suburban cases alike. In \cite{lina_wu}, the authors saw a noticeable difference in the accuracy of their ANN, when they included environmental features as part of their input. These environmental features corresponded to different elements of a suburban terrain, such as roads, tunnels and buildings. The length of the straight line connecting the receiver and the transmitter within each of these was used as an input feature. In \cite{sotiroudis_3}, the authors found that even though global input information about an RT-generated urban grid was more important than local one, they could replace it with a reduced input representation that led to more accurate predictions. That input representation consisted of the path profile between the receiver and the transmitter, as well as local information about the receiver. Uncorrelated input features have a bigger impact on the performance of the ML model. In \cite{ferreira}, the authors achieved no substantial improvement in their ANN model of an urban scenario, when they added the signal strength computed with a knife-edge diffraction model. That extra input feature though was highly correlated with one of the initial inputs to the ANN, that of the diffraction loss computed by the same method.

\subsection{Training data}

\subsubsection{Size of training dataset}

Increasing the volume of training data is always helpful, as long as the ML model is not in the overfitting regime. That is because the model can explore a bigger space of {$V_x$} as part of its training. Hence, its predictions can be more reliable. This has been observed in several papers \cite{ferreira, moreta}. The new training data have to be as representative as possible of the propagation problem. For example, in \cite{anitzine}, the authors used two different sized training sets. The smaller dataset contained a small number of cases where the signal reached the receiver after reflecting off from buildings' walls. The bigger training set gave more accurate predictions, not only because it contained more training data in general, but also because it included a wider collection of multipath propagation cases. We should note that the ML model has to be complex enough to accommodate the increased training set, otherwise, underfitting will occur. The same also applies when increasing the number of input features.

In classification problems, extra consideration has to be given when constructing the training set. The number of samples collected from each class distribution should be comparable to avoid bias. \cite{Krawczyk}. As an example, in \cite{jang}, the authors investigated how a balanced training set could impact the accuracy of an ML classifier, predicting the building and floor a user is located at. Balancing the initial dataset improved the localization accuracy by about 1\%.


\subsubsection{Dataset augmentation}

Increasing the volume of measured training data is an expensive and time-consuming process. In these cases, data augmentation techniques can be used. When there is a visual input to the ML model in the form of images, various image transformations can be applied to create new images that can be used as new training data. For example, in \cite{lee}, every input image was rotated by 1\textdegree, achieving an increase of the training set by a factor of 180. In \cite{thrane}, apart from rotation, the authors also used sheering for the training images. The new images were indeed correlated to the old ones, however, they could still be useful since the data demands of any ML model are considerable. 

Simulated data may also be used to increase the training set. In \cite{wen}, the authors used training samples coming from a log-distance model to improve the ML model predictions in a newly employed frequency, in an airplane cabin. To train their model, they only used those coordinate points inside the aircraft where the log-distance model showed good agreement with the actual measured values. When they tested their model on a new frequency, the authors found that the accuracy improved, and was in fact higher than by just using an ML model trained without these points. However, when the authors included a small part of measurements at the new frequency band, the accuracy improved even more. Nevertheless, even using seemingly lower-quality data (data coming from an empirical model), can be helpful during the training of the ML model. That was also validated when the authors used a fusion of measured and empirical data in their final experiment, improving the accuracy of their ML model even further.

Fusing training data from various sources may be necessary to construct large training datasets. It has also been shown to improve the performance of the ML model. As an example, in \cite{adege}, the authors used training data coming from UAV, Wi-Fi and cellular base stations. In \cite{yadav}, the authors used a fusion of Wi-Fi RSS and geomagnetic field (GMF) data. The authors also found that by using only RSS or GMF data, the accuracy of their ML model deteriorated. Likewise, in \cite{gan}, the authors trained their model using a fusion of measured  (Wi-Fi, Bluetooth, GM data) as well as synthetic data (RT).  


Finally, generative models can also be used to produce additional training data. In \cite{li}, the authors used a generative adversarial network (GAN) to boost their measured dataset by generating additional channel state information (CSI, \cite{molisch}) amplitude maps of an indoor environment. They found that the classification accuracy of their network significantly improved. Likewise, in \cite{belmonte}, the authors used GANs for data recovery. Their goal was user tracking, done by measuring successive locations of a moving user. In cases where no measurements were collected (eg. not reachable from an access point (AP)), the GAN was used to estimate the user's location.


\subsubsection{Impact of training data}

Just like uncorrelated input features may improve the accuracy of an ML model more than correlated ones, uncorrelated training samples are also more useful than correlated ones. As an example, in \cite{wen}, the authors decided to use 20\% of their measured data to train their ML models for a new frequency deployment in the airplane cabin previously mentioned. Their experiments showed that using measurements taken uniformly across the cabin rows led to a more accurate model, than using measurements taken only at the front or back rows of the airplane. The geometry at the front of the plane differed from the one at the back, therefore, taking the majority of measurements at either of these areas led to a non-representative of the airplane's geometry dataset. Finding similarities among the training data is helpful and can lead to a more balanced and representative training set. Clustering algorithms can be used to group the training data and present the ML model with evenly distributed training samples among the clusters. Such a procedure was followed in \cite{moreta} to cluster the coordinates (longitude, latitude) and the altitude of the training samples, using the well-known $k$-means clustering algorithm. However, this procedure added another hyperparameter to the ML propagation modeling process, that of determining the number of clusters $k$.

\subsubsection{Type of training data}

As already discussed, ML models can be very demanding in terms of training data, often making the task of collecting large sets of measured data infeasible. Instead of measurements, synthetic data generated by electromagnetic solvers may also be used as training data for ML propagation models. For these cases, there is an additional choice to be made, that of the solver. Some of them may be computationally more efficient than others. Moreover, it may be easier to construct the simulation environment in some solvers than others. One of the most popular ones for outdoor environments is RT \cite{sotiroudis_1, sotiroudis_2, cerri, kuno, ahmadien, falcone}. Another deterministic method used in propagation modeling simulations is the VPE method. The method assumes a paraxial approximation with respect to the direction of propagation of the wave. Therefore, it is often used in simulating enclosed environments that have waveguiding characteristics \cite{aris}, or terrestrial propagation scenarios \cite{cheng1}.

The use of physics-based solvers for generating the training/test data also leads to input features that are usually different from the ones used in measurement-based training of ML models. These input features are solver-specific. For example, in \cite{cerri}, the authors included parameters such as the number of reflected and diffracted rays that reach the receiver according to their RT. Grid-based methods, such as VPE, allow for the assignment of input features on individual “cells”. Likewise, RT employs reflecting surfaces, whose specification introduces input features for the model. To that end, extra input parameters may be used to convey information relating to each cell of the grid. In \cite{sotiroudis_2}, the authors designed a grid representing an urban environment. In \cite{kuno}, a public square in front of a station was implemented into RT. In both cases, cell-specific information was provided to the ML model as input. In \cite{sotiroudis_2}, it was a parameter indicating whether the cell was indoor or outdoor. In \cite{kuno}, it was the maximum obstacle height within the cell.


\begin{figure*}[h!]\centering
\vspace*{-0.4cm}
\begin{tikzpicture}
\node at(0,0){\includegraphics[scale=0.62,clip,trim={0cm 0cm 0cm 0cm}]{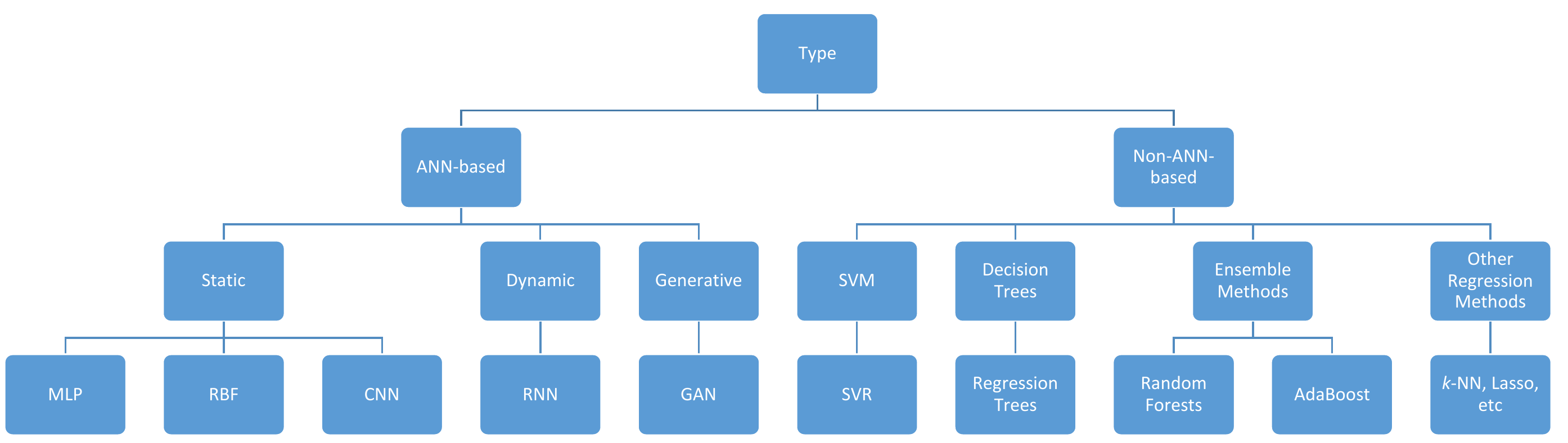}};

\end{tikzpicture}
\caption{Diagram of the main ML models used for radio propagation.}
\label{fig:ml_models}
\vspace*{-0.4cm}
\end{figure*}

\section{ML model}

As already mentioned, one important decision for the accuracy of the propagation model is what type of ML method to use. This section separates the propagation papers into three main groups. The number of propagation modeling papers that have used ANNs is significant. We expect this trend to continue due to the growing research activity on ANNs. Thus, the first two groups are defined by whether the ML model used is ANN or non-ANN based. The last group describes hybrid models that combine more than one ML methods. A schematic diagram on the main ML models that this paper considers can be seen in Fig.~\ref{fig:ml_models}. Finally, we discuss the constraints in the range of the input features, and how these can lead to relatively constrained or more general ML propagation models.

There are many different ML models. More complex models may perform better than others, when given large datasets or a large number of input features. Simpler models may be faster for small-scale tasks. For any ML model, its hyperparameters play an important role. These can pertain to the architecture of the ML model or can be directly connected to its training. The number of hidden layers and nodes for an ANN is an example of the first category, while hyperparameters such as the learning algorithm, the amount of the learning rate or the type of kernel functions for the SVMs are part of the second category. The values of the hyperparameters can be set via several heuristic techniques, such as grid search, random search and others \cite{hyper}.

\subsection{ANN-based models}
This subsection covers papers that utilize both the standard multi-layer perceptron (MLP) ANNs as well as their many variations, such as RBF-ANNs, convolutional neural networks (CNNs), recurrent neural networks (RNNs) and GANs.

Generally, adding depth to an ANN by increasing the number of layers, or increasing the number of neurons/nodes, improves its accuracy. This has been reported in various papers \cite{ostlin, lina_wu, lee, jang}, where the authors experimented with different numbers of layers and found that deeper ANNs gave more accurate predictions. Similar findings were reported in \cite{chang} and \cite{ferreira}, where the authors observed that increasing the number of neurons while keeping the number of hidden layers constant, improved the accuracy of the RBF-ANN. Bigger ANNs (and generally more complex models) are more prone to overfitting, especially if only limited numbers of training data are available. Hence, regularization techniques have to be used, such as early stopping or L1 and L2 regularization \cite{ml_book, goodfellow}. The former method stops network training when the validation error follows an increasing trend, a sign of overfitting. The latter penalize big network weights by incorporating the L1 and L2 weight norm, respectively, into the cost function (see Eq.~(\ref{eq:cost_function}) in the Appendix).

Many variations of MLP-ANNs have been developed over the years, with some of them utilized for propagation modeling. One variation of the standard MLP-ANN, widely employed in propagation modeling, is the RBF-ANN \cite{chang, popescu_1}. For these models, the choice of radial basis functions is just as important as choosing an efficient activation function for the MLP-ANNs. In \cite{chang}, the authors used spline functions with good results. Other types of ANNs have also been used such as the wavelet neural networks (WNNs) \cite{cheng2} that use a wavelet as an activation function.


A variation of ANNs, highly popular for classification tasks and widely used for image recognition, is the CNNs \cite{goodfellow}. The input to a CNN is typically a 3-dimensional tensor, instead of input vectors used in the standard ANN.  For example, for image classification tasks, each input layer of the data contains pixel values for each of the three RGB color channels in the form of a 2-dimensional matrix. Stacking these three matrices together will form the 3-dimensional input tensor. The same procedure is followed when grey-scale images are used as input \cite{thrane, ahmadien}. In the context of propagation modeling, the input to the CNN does not have to be visual, i.e. instead of color intensities, each pixel can represent other useful information. Moreover, a varying number of input layers/channels can be used. For example, in \cite{lee}, the authors used two input channels. One, encoded the normalized height of each building in an RT-generated city grid. The other, accounted for the normalized difference in height between the transmitter and the ground level at each point of the grid. In \cite{kuno}, the authors created a square grid representing an urban square that was  modeled in RT. They used three channels containing information regarding the distance from the transmitter, the distance from the receiver, as well as the height of each cell on the grid, respectively. Three layers of input features were also used in \cite{imai}. Furthermore, the input is not required to contain spatial information. For example, in \cite{jang}, the authors used images of RSS coming from multiple APs for a given indoor location. Instead of using an input vector containing these values, they converted them into a 2-dimensional format by zero-padding. Some of the most popular CNN implementations in computer vision have also been used for propagation modeling. Among them are VGG-16 \cite{vgg} as well residual network (ResNets) implementations \cite{resnet}. CNNs use filters (feature detectors) to learn mappings between the input and output space. A parameter sharing scheme exists among filters, i.e. the same filter is applied at different parts of the input image \cite{goodfellow}. That trait as well as sparsity of connections make CNNs more efficient than standard MLPs \cite{goodfellow}.



Another class of ANNs that has recently grown in popularity and utilized in propagation modeling, is the RNN. Contrary to the previously discussed ANN types, these networks exhibit a dynamic behaviour, whereby their output depends not only on the current input, but on previous inputs as well. Hence, RNNs process sequences of input data in a recurrent fashion. Many different variations of RNNs have been used for propagation modeling purposes, such as the echo state networks (ESNs), \cite{gideon}, Elman RNNs \cite{cheng2}, standard RNNs \cite{turabieh, hoang} and gated RNNs that use a gated recurrent unit (GRU) instead of a standard recurrent unit cell \cite{adege, hoang}. The most important and popular type of RNN is the long short-term memory (LSTM) RNN that can capture longer dependencies in the input data, compared to the other types of RNNs \cite{xu, wang, hoang, yadav, hsieh}.  In propagation modeling problems, the input sequence to the RNN is usually spatial \cite{hoang} or temporal \cite{adege, turabieh}. The ability to learn temporal sequences also makes RNNs ideal for modeling time variations in a communication channel.


Finally, another recent and very powerful class of ML models involves GANs. \cite{goodfellow_gan}. These models consist of two neural networks, usually CNNs \cite{dcgan}, the generator (G) and the discriminator (D). G tries to model the distribution of the real (target) data, eg. an image of a human face, by producing fake images, while D tries to discriminate between these and the real images that are provided to it. Based on this adversarial game, a trained G can produce artificial/synthetic data that are almost indistinguishable from the real ones. Whereas the inception of GANs envisaged them in an unsupervised setting, they have been adapted accordingly for semi-supervised learning by providing D with labeled data. Moreover, a useful type of GANs for propagation modeling is conditional GANs (cGANs) \cite{cgan}. These can include constraints on the data produced by G. GANs' generative ability is highly utilized in propagation modeling for dataset augmentation \cite{belmonte, li} or to improve the reliability of ML models when the number of labeled data is small \cite{chen}. Finally, apart from GANs, other generative networks can also be used, such as deep belief networks (DBN) \cite{gan}.

\subsection{Non-ANN-based models}

Apart from ANNs, many other ML models can be used for regression tasks, such as computing PL in a given communication link. Even though they differ from ANNs in their tuning parameters, architecture or learning algorithm, the main challenge remains the same; namely, what network parameters to choose for an efficient configuration.

One popular choice outside ANNs is SVMs and more specifically their regression (SVR) version \cite{piacentini, timoteo, gideon, wen, moreta}. SVM kernels are equivalent to the activation functions of the ANNs. The type of kernel is important for the accuracy of the model. In \cite{timoteo}, the authors experimented with three different kernels, namely a polynomial, a Laplacian and a Gaussian kernel. Results on the test set showed that the Laplacian kernel gave more accurate predictions than the other two. Other kernels, such as RBF kernels, have also been used for propagation modeling \cite{moreta}.

Another alternative is a genetic algorithm (GA), used to derive a closed form expression of the received PL \cite{fernandes}. Other well known regression algorithms, such as the $k$-nearest neighbors, are also popular for radio propagation modeling \cite{oroza, moreta}. Recently, ensemble methods have been used with promising results \cite{karra}. Their function is based on the simple notion that a combination of learners can be more powerful than each one separately. Random forests (RF) \cite{ensemble_book}, which operate by constructing a variety of decision trees, is such an ensemble method, used for radio propagation modeling \cite{karra, oroza, wen, moreta}. Another form of ensemble learning is boosting. One of its more popular implementations, the Adaboost algorithm \cite{ensemble_book}, has also been applied for propagation modeling \cite{karra, oroza, wen, moreta}. Any one of the models previously mentioned, such as the SVR or the $k$-NN algorithms, can be used as a learner for the ensemble method. Since ensemble methods require the participation of many learners for the generation of the final output, a weighting scheme between them has to be chosen. Finally, it should be noted that training ensemble methods may take considerable time, compared to separately training the base learner.

\subsection{Hybrid models}

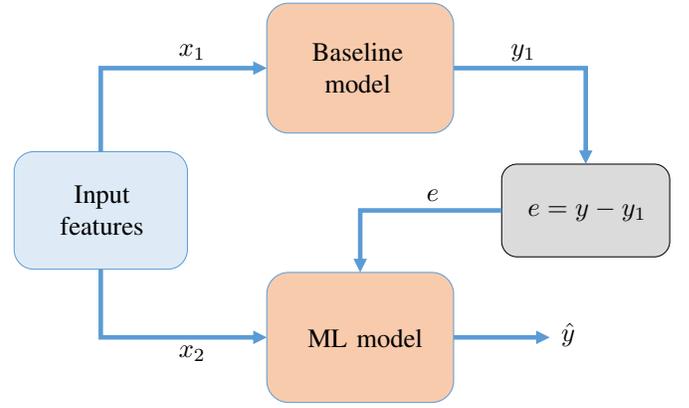
\begin{figure}[!htb]\centering
\begin{tikzpicture}
\vspace*{-0.2cm}
\node at(0,0){\includegraphics[scale=0.38,clip,trim=
{0cm 0cm 0cm 0cm}]{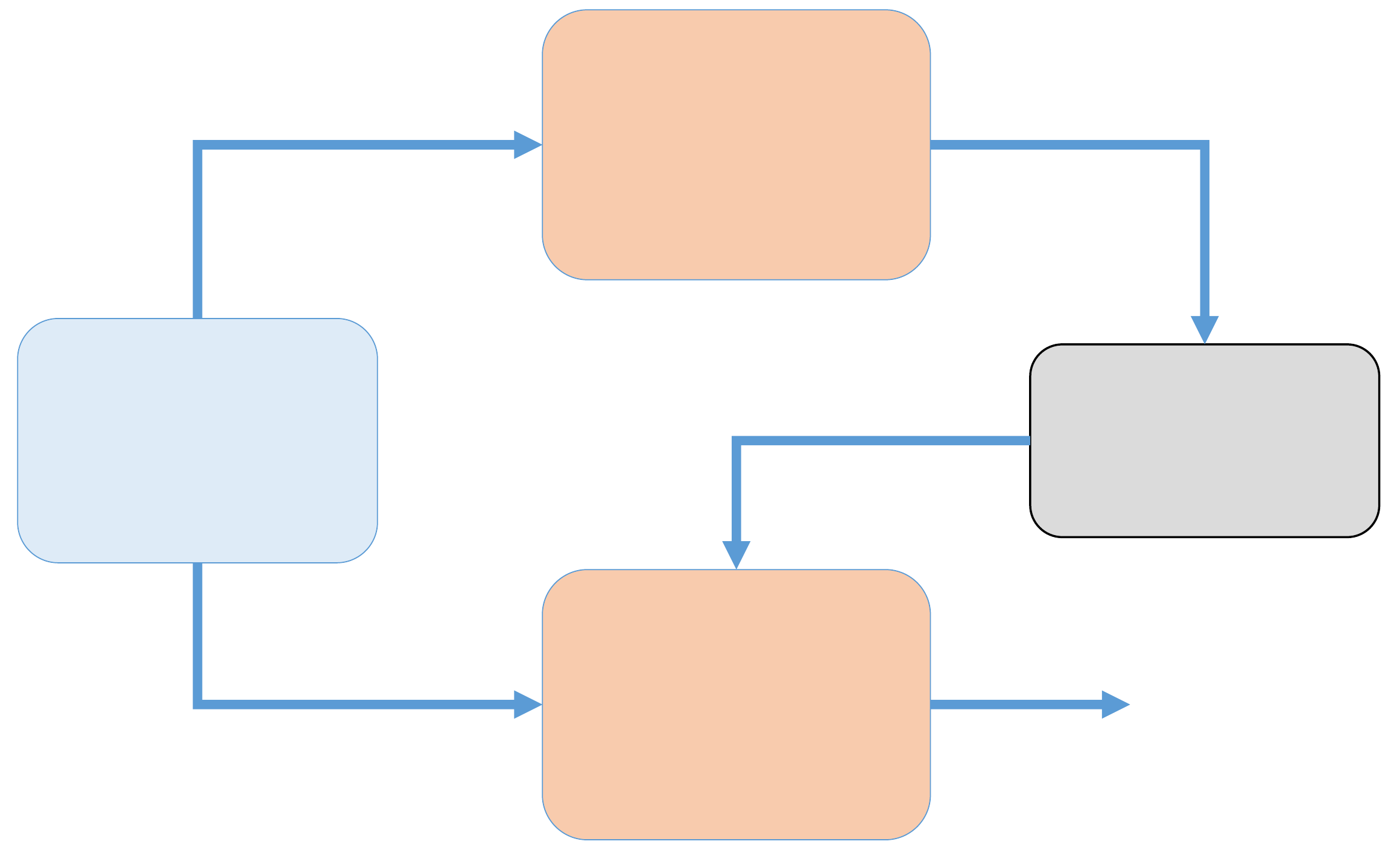}};

\node at(-3.2, -0.1) [align=center] {Input\\ features};

\node at(0.2, 1.8) [align=center] {Baseline \\ model};

\node at(0.3,-1.8) {ML model};

\node at(-2, 2) {$x_1$};
\node at(-2, -2) {$x_2$};

\node at(2.4, 2) {$y_1$};

\node at(1.2, 0.1) {$e$};

\node at(3.25, -0.1) {$e=y-y_1$};

\node at(3, -1.75) {$\hat{y}$};

\end{tikzpicture}
\caption{The architecture of an error correction hybrid model.}
\label{fig:hybrid}
\vspace*{-0.5cm}
\end{figure}

Instead of creating an ML model to directly predict various propagation parameters, one can implement a correction mechanism for an existing propagation model. The main goal of the ML model is to enhance the knowledge provided by a baseline propagation model, by either ``learning" its errors or by using its predictions as part of the ML model's input. This combination of a baseline and an ML model can be considered as a hybrid approach. The architecture of one such error correction hybrid model can be seen in Fig.~\ref{fig:hybrid}. The input to the baseline and the ML model can be generally different. Considering one sample point, the prediction of the baseline model $y_1$ is used to compute the error $e$ with respect to the real target value $y$. Then, this error is used as the target value for the ML model. Therefore, the prediction of the ML model $\hat{y}$ will correspond to that learned error. Consequently, the final prediction of the hybrid model would consist of the sum of $y_1$ and $\hat{y}$. The initial prediction of the baseline model can be thought of as a starting point, from which the ML model can improve on the baseline model's predictions. Hence, the training of the hybrid model is expected to be more efficient.

Various papers over the past years have used ML models as error correction tools in predicting various propagation parameters of a communication channel. The baseline models that have been used in the literature are mostly empirical ones. For example, in \cite{popescu_1, gschwendtner, cheerla}, the COST-Walfisch-Ikegami (CWI) model was used as a baseline model. In \cite{isabona}, a log-distance model was used to drive the training of the ML model, while in \cite{panda}, Hata's model provided the starting knowledge. On the other hand, the baseline model can be an ML model itself. For example, in \cite{ehbota}, the base model consisted of an Adaline (adaptive linear network) NN, a very basic network computing only the linear weighted sum of its input. The input to the ML model can be the same as the one of the baseline model \cite{popescu_1, ehbota}. It can also include input parameters that are not modeled by the baseline model itself \cite{gschwendtner}. Finally, the ML model can also improve on the predictions of a lower fidelity model, to match those of a higher one. For example, in \cite{falcone}, an ANN was used to improve on the RSS predictions of an RT using 4 reflections. The authors found its accuracy was close to the case of using 7 reflections, but its runtime smaller.

Hybridization can also refer to the interconnection of several ML models to create a more complex one. Such a procedure was followed in \cite{thrane}. The authors used two ANNs and a single CNN to synthesize a more complex ML model. The task of the CNN was to learn latent features from digital images, while the first ANN was tasked with learning mappings from input features pertaining to the transmitter and receiver, as well as their distance. Finally, the second ANN was used to concatenate the outputs of the other two models and produce the final output. Likewise, in \cite{imai}, the authors first used a CNN to extract latent features from RT-generated maps of urban environments. Those features along with the inputs of the CNN were then used as input to a 2-hidden-layer ANN, which in turn generated the output (PL) of the cascaded network. Another cascaded-layered network was created in \cite{turabieh}. The authors combined two RNNs, with the first classifying the building where the user may be located, and the second the corresponding room. Finally, in \cite{wang}, the authors used a hybrid of a ResNet and an LSTM network. ResNet learned the spatial correlation between RSS samples at a given 
timestamp in the form of black-and-white radio maps. Sequences of 5 RSS maps were then stacked together and passed on to the LSTM to extract the temporal correlations between them.


GANs can be thought of as the interconnection of two baseline models (eg. CNNs). However, some papers hybridized GANs with other models to perform even more complex functions. For example, in \cite{li}, the authors interconnected a DCGAN with a CNN for classifying user's location. In \cite{chen}, the authors used a DCGAN employing a CNN classifier sharing weights with D. In that case, both D and the classifier learned concurrently as a dual separate model, with the first classifying user location in an indoor environment, and the latter discriminating between real and fake samples. In \cite{belmonte}, the authors used a cGAN in combination with an LSTM network. The RSS maps of the modeling environment created by the cGAN were used by the LSTM for user tracking reasons. For the same purpose, the authors in \cite{gan}, combined an unsupervised DBN with a supervised classifier.


As can be seen from Fig.~\ref{fig:hybrid}, the operation of the baseline and the ML model are linked together. However, they still exist as two separate models. One interesting alternative is to merge the two models and integrate the baseline model into the ML model itself, as was the case in \cite{panda}. One of the ANN nodes was used to implement the Hata model. The output of that node was connected to the total output of the network through a unit weight, to ensure that the final output would be always influenced by the baseline model's predictions.

\subsection{Input-constrained models}

A propagation model can consist of separate ML-based models derived for a subset of the frequency range of interest, specific cases of the receiver position (e.g. LOS or NLOS locations), or other subsets for specific cases of the input features. For example, several authors have created frequency- \cite{moreta}, environment- \cite{anitzine}, route- \cite{ferreira, anitzine}, or distance-specific \cite{sotiroudis_3} ML models, respectively. Nevertheless, more general ML propagation models can be created too. One such example of designing a multi-frequency, multi-environment propagation model can be found in \cite{ayadi}. To that end, the authors collected measurements at a variety of different frequency bands and at areas ranging from urban to rural. More examples of multi-frequency networks can be found in \cite{cheng1, cheerla, wen}. A similar procedure was followed in \cite{oroza}, where the forested areas measured had different features, such as canopy density and terrain complexity. Their ML model accounted for all these diverse environments. Relaxing the constraints on the range of the input features requires the training of more complex ML models that demand bigger volumes of training data and higher computing power. On the other hand, since they are general, they are more flexible.


\section{Output}

In this section, we present different types of outputs that have been considered in ML-based propagation models. We also discuss the accuracy of these models and connect some output errors to specific input features. As a note, for regression problems, various error metrics can been used to evaluate the accuracy of the ML model on the test set. The mean absolute error (MAE), the mean squared error (MSE) or the correlation factor (CF) between the model predictions and the target values are just some of these. For classification tasks, the accuracy of the model corresponds to the probability of correctly predicting the output class.

\subsection{Type of output}

Even though most ML-based propagation scenarios correspond to regression tasks, such as those previously described in the paper, classification problems have also been considered. As an example, in \cite{zhang_y}, the authors used high-level information at the receiver, such as PLE, RMS delay, RMS angular spread and others, about various measured as well as simulated communication scenarios. Then, they used these as inputs to classify each environment into urban or rural, each one having also an LOS or NLOS specification. In some classification problems, choosing the number of different classes may not be completely straightforward and the labeling may also be subject to human errors. In \cite{zhang_y}, an environment could have both urban and rural features, therefore, its classification as urban or rural may be misleading. In such cases, unsupervised algorithms may be used to cluster the data. In \cite{zhang_y}, the authors experimented with 2 supervised algorithms, $k$-NN and SVM, as well as with 2 unsupervised, the $k$-means \cite{hastie} and a probabilistic inference model, the Gaussian mixture model (GMM). 
It is interesting to note that the authors found that the optimal number of clusters for the $k$-means algorithm was equal to the number of classifying classes.

As already mentioned, the output of an ML model can
also be probabilistic. In \cite{adeogun}, the authors trained an ANN using channel statistics (moments and covariances of channel transfer functions) over many simulated realisations of two stochastic propagation models; the path graph (PGr) and the Saleh-Valenzuela (SV) model. The trained ANN was able to generate
valid parameters for the two probabilistic models, so that both showed a similar power density profile to the actual one. Likewise, in \cite{ahmadien}, the authors used a CNN-based approach for PL distribution estimation in a variety of different urban and suburban environments. Satellite images of the environments were first converted to 3D models that were imported into an RT solver. Then, the simulated PL was used to generate the PL probability density function in each environment. In such cases, the number of bins that comprise the distributions should be carefully chosen.

Similar to the sequential input of an RNN, its output can be sequential too. In \cite{hoang}, the authors investigated whether their multiple-input multiple-output (MIMO) LSTM was more accurate than a multiple-input single-output (MISO) one. They found that the MIMO model was more accurate. Similar findings were reported in \cite{belmonte}, where a MIMO model was more accurate than a single-input single-output (SISO) LSTM.

An ML propagation model that learns the RSS inside a specific geometry can be used to compute other parameters and influence network-level decisions. As an example, in \cite{monteiro}, the authors created an ANN that outputs the conductivity and electric permittivity of the ground. Moreover, large-scale fading outputs, such as the PLE of a communication  channel, are also helpful for a higher-level analysis of a given environment \cite{lee}. ML propagation models that learn RSS maps can also be used for localization and tracking, especially in indoor environments. Examples include models that learn to estimate user location, both as a regression \cite{jang, turabieh} as well as a classification problem \cite{adege, xu}. The former is done by computing the coordinates of the user, while the latter by classifying the subspace the user is located in.

\subsection{Understanding output errors}

Some output errors of ML-based propagation models can be attributed to certain input features. For example, the more complex the multipath mechanisms are in a given environment, the more error-prone becomes an ML model of the channel impulse response of a communication system in that environment. Similar connections are made in the following.

\subsubsection{Type of environment}

The type of environment, as well as the presence or absence of an LOS component, are two factors that influence the error of ML models. ML propagation models in urban scenarios exhibit lower accuracy than in rural \cite{ayadi} or semi-urban \cite{popescu_1, ahmadien} environments. A similar finding was reported in \cite{lee}. The authors noted a slight increase in error when they inserted more buildings in their RT-generated urban grid, i.e. when they made the urban environment even more complex. Moreover, ML models for NLOS scenarios are also less accurate than LOS ones \cite{popescu_2, kuno, anitzine, xu}. Similar results were reported in \cite{zhang_y}, where the authors trained and tested their network using two different routes in an urban environment and ascertained that the route that maintained better LOS conditions exhibited smaller prediction errors. Also in \cite{sotiroudis_3}, the largest errors corresponded to receiving points for which LOS was obstructed by several tall buildings. For such cases, the authors found that local (around the receiver) information could improve accuracy. In addition to that, increasing the height of the transmitter had the same effect, since by doing so, the signal strength at the same receiving points increased (some NLOS cases also changed into LOS ones). Similar findings with respect to the transmitter antenna height were reported in \cite{ahmadien}.

\subsubsection{Distance and frequency}

Another factor affecting the accuracy of predictions is the simulated distance. Generally, errors are higher when the receiver is in the vicinity of the transmitter. For example, in \cite{ayadi}, the authors concluded that for distances smaller that 500\,m, the ANN presented higher errors than for larger distances. Similar findings were reported in \cite{lee}. A workaround is to build distance-specific ML models. In \cite{sotiroudis_3}, the authors created three distance-specific ANNs, for propagation distances less than 350\,m, between 350\,m and 700\,m, and larger than 700 m. Interestingly, the ANN trained on small distances was slightly more accurate than the one trained on medium distances, due to overtraining of the first network on learning the more pronounced fast fading characteristics of smaller distances. When that network was tested on medium or large distances, it was found less accurate than the network trained on medium distances. For a distance-agnostic ML model, the higher errors observed closer to the transmitter are related to the break-point of the communication channel \cite{bertoni} and the near field of the transmitting antenna, both of which depend on the frequency of operation. Generally, as frequency increases, the small-scale fading characteristics of the channel become more pronounced. That leads to increasing errors for the ML model \cite{cheerla, jo, aris, ahmadien}. For example, in \cite{wen}, the authors observed increased errors at the frequency band of 3.52\,GHz and  5.8\,GHz compared to the one of 2.4\,GHz. This is also demonstrated in the case study of the Appendix. On the other hand, in \cite{ayadi}, the authors concluded that their ANN's errors did not follow a specific pattern across the frequency range used (450\,MHz\,-\,2600\,MHz). However, they averaged their measurements over a 40-wavelength distance, reducing the effects of fast fading.

\subsection{Accuracy of ML models}

\subsubsection{Comparison to empirical models}

In early ML propagation papers, empirical models were mainly used as a reference for the accuracy of the ML models. Many papers have showcased the higher accuracy of ML propagation models over various empirical ones. In \cite{popescu_2}, the authors showed that the ANNs they created, both for LOS and NLOS cases, outperformed three reference empirical models used, namely the Walfish-Bertoni (WB) model, the single-slope model and the CWI model. In \cite{cheerla}, the MLP-ANN implementation was more accurate than the ARMA forecasting model at both frequencies of 800 and 1800\,MHz, while in \cite{ostlin}, the ANN was found more accurate than Recommendation ITU-R P.1546 for rural areas and the Hata model. Finally, in \cite{ferreira}, the ANNs outperformed empirical models that also included diffraction losses, such as the Cascade Knife Edge (CKE) and Delta-Bullington model. Similar observations were made for other ML models. As a matter of fact, in \cite{fraile}, the RBF-ANN implementation was found more accurate than Meeks \cite{meeks} and Maximum Shadowing (simplified Meeks) empirical models. In \cite{popescu_1}, the same type of network was found more accurate than the single-slope, WB and CWI models. Similar results were also observed in papers using hybrid ML models, such as in \cite{gschwendtner, popescu_1, thrane}. In \cite{panda}, the authors also showed that a combined Hata-ANN model was more accurate than each one of its constituent models.


\subsubsection{Comparison to other ML models}


Other papers compared different ML models to identify the most accurate for their application. In \cite{timoteo}, the authors found that results of the MLP-ANN were very close to that of an SVR. Similar findings were reported in \cite{popescu_1, piacentini}, where the accuracy of the RBF compared to the MLP-ANN was only marginally higher. In \cite{cheng1}, the authors concluded that for their setup, the WNN implementation was more accurate than the RBF-ANN. In \cite{gideon}, the authors found that the ESN was more accurate than an SVR implementation, but also more time consuming during training. In \cite{ribero}, the CNN outperformed a standard ANN for predicting PG over irregular terrain. 
In \cite{jang}, the authors also found that their CNN was more accurate than an MLP-ANN classifier and a stacked auto-encoder (SAE). RNNs have also been found more accurate and faster than MLP-ANNs in the modeling of time-varying communication channels \cite{adege, hoang, xu}. Among the RNN models, the authors in \cite{adege} found that GRU outperformed their LSTM implementation. On the other hand, LSTMs were found more accurate than both standard RNNs \cite{hoang, hsieh}, as well as GRUs \cite{hoang}. 



Regarding ensemble implementations, in \cite{oroza}, their RF performed more accurately than the MLP-ANN, $k$-NN and AdaBoost implementations. Similar results were observed in \cite{karra}, where the author's RF implementation outperformed the $k$-NN, SVR, Adaboost and the gradient tree boosting (GTB) methods. However, it should be noted that all 5 ML models were close, accuracy-wise. Finally, a voting regressor (VR) ensemble model constructed from these 5 learners was slightly more accurate. One more paper that found RF to be the most accurate model is \cite{wen}, where the authors compared it against an empirical model (log-distance) and 3 other ML models; a single hidden-layer ANN, SVR and AdaBoost. RF was the most accurate model across all three frequencies the authors used (2.4, 3.52 and 5.8\,GHz). In \cite{zhang_y}, the RF implementation was again found more accurate than the MLP-ANN and the SVR. Finally, in \cite{moreta}, the authors compared AdaBoost, RF, $k$-NN, SVR and ANN regression algorithms against Lasso regression \cite{lasso} (a type of linear restricted regression) and the kriging algorithm 
(often used in geostatistics). The goal was to predict the RSS of a digital terrestrial television (DTT) system at any point inside a coverage area and project it on Google maps. They found that all ML models were noticeably more accurate than kriging and Lasso. Among the ML models, the most accurate were AdaBoost and the RF regression algorithm, exhibiting similar performance.

\subsection{Generalizability and test set}

As discussed in Section II, the ML model should have good generalization abilities. This is determined by how accurate the ML model is when evaluated on the test set.

\subsubsection{Composition of the test set}

For any reliable conclusion about the generalizability of an ML model, the test set should contain data samples that have not been used during training. The kind of samples is implementation-specific and connected to the goal of the ML model. In \cite{neskovic}, the authors tested their ANN using a transmitter location that was different from the one used to train it. In \cite{ayadi} and \cite{ferreira}, the authors used different routes to test and train their ANN. Likewise, for cases where synthetic data were used, as in \cite{sotiroudis_2}, the authors trained their network on a grid of uniformly placed streets and buildings, but tested it on a non-uniform grid. The test samples can represent an interpolation problem, i.e. when the values of the test features fall within the range of the training inputs, or they can be extrapolation samples. For example, in \cite{cheerla}, the generalization abilities of the ML model were tested for receiver distances higher 
than those used during the training, simulating a time-series forecasting problem. Moreover, in \cite{wen}, the authors tested their network on frequencies outside the ones used to train it.
Another extrapolation problem can be found in \cite{cheng2}, where the authors used an Elman RNN to calculate the propagation factor in VPE simulations.

When an ML model is trained only on synthetic data, its accuracy is bounded by the accuracy of the solver that generates the data. Therefore, the solver itself has to be accurate. In order to enhance the reliability and thus the generalizability of such an ML model, measurements can also be used as test cases. We presented such an example in \cite{aris}, where we used a fusion of synthetic as well as measured data to evaluate an ML model. More specifically, an ANN was trained on RT-generated data simulating various arch-shaped tunnels of different cross-sections. Afterwards, the ANN was tested not only on RT-generated test tunnel configurations, but also on sections of the London Underground.

\subsubsection{Test set distribution}

As explained in Section III, the training set has to be representative of the test set. Ideally, both datasets should be generated by the same distribution. Otherwise, large errors may be observed. As an example, in \cite{ferreira}, the authors created two different ANNs for two measured routes. Both ANNs were highly accurate when tested on the routes that were used to train them. However, when measured data from one route was used to generate predictions for the other, the performance of the ML model deteriorated. In fact, accuracy was lower when using data from route 1 to generate predictions for route 2, than the other way around. The reason for that was that route 1 crossed an urban area, while route 2 crossed a mix of urban, rural and semi-rural terrain. Hence, it included a wider range of propagation effects than route 1. As another example, the authors in \cite{sotiroudis_2} trained their ML model on RT-generated data inside a uniform urban grid. When they tested their model on data generated in a non-uniform grid, they noticed a considerable increase in the prediction error.

\section{Conclusion}

In this paper, we gave an overview of ML-driven propagation models. We described what an ML propagation model is and how it can be constructed. We analyzed its three building blocks: its input, the ML model used and its output. Moreover, we presented the challenges associated with each one of these blocks and how they were tackled by several relevant papers in the literature. This was done in a systematic way, by focusing on the methods, conclusions and higher-level decisions of each paper. More specifically, we can conclude that:

\begin{itemize}

\item Input features should convey useful information about the propagation problem at hand, while also having small correlation between them.

\item Dimensionality reduction techniques can help identifying the dominant propagation-related input features by removing redundant ones. 

\item Increasing the number of training data by presenting the ML model with more propagation scenarios improves its accuracy.

\item Synthetic data generated by high-fidelity solvers, such as RT or VPE, or empirical propagation models, can be used to increase the size of the training set and refine the accuracy of ML based models.. Data augmentation techniques can also be used for that purpose.

\item Regarding the accuracy of the ML models, RF was found to be the most accurate by a number of papers. Generally though, the differences in accuracy between the various ML models are implementation-dependant and were not large for the ML models we reviewed.

\item More general ML propagation models, covering a wide range of frequencies and propagation environments, require more training data than simpler ones. The same applies for models that correspond to more complex propagation scenarios, such as in urban environments.

\item ML models can be connected to create hybrid ones that can be employed in more complex propagation problems. 

\item The evaluation of an ML model for a given propagation problem requires a test set modeling all present propagation mechanisms. Its samples should come from the same distribution as that of the training samples.

\end{itemize}

There are still open problems or questions to be further investigated in this area. Examples of such problems are: 
\begin{itemize}

\item The connection of the physical propagation mechanisms present in a channel with the architecture and the volume of training data required for the development of an ML-based channel model.

\item The potential of ML methods to simplify the input required for the development of PL models in complex environments (e.g. replacing elaborate CAD models with a sequence of images of the environment to be processed by the model).

\item Further research on GANs and how they can improve the accuracy of an ML propagation model, especially in the regime of low-volume training data.


\item Reinforcement learning (RL \cite{sutton}) (being the third class of ML-related problems, apart from supervised and unsupervised ones) has been used for wireless channel modeling, tackling tasks such as interference mitigation \cite{rl1} or resource and channel allocation \cite{rl2}. Investigation of RL techniques for electromagnetic wave propagation modeling seems highly promising.

\end{itemize}

We do believe though that advances in the ML field will keep influencing and advancing the area of radio propagation modeling.

\section*{Appendix - a case study of an ml propagation model}\label{sec_appendix}

\begin{figure}
\centering
\vspace*{-0.3cm}
\begin{tikzpicture}
\node (image) (0,0) {\includegraphics[scale=0.45,clip,trim={0cm 0cm 0cm 0cm}]{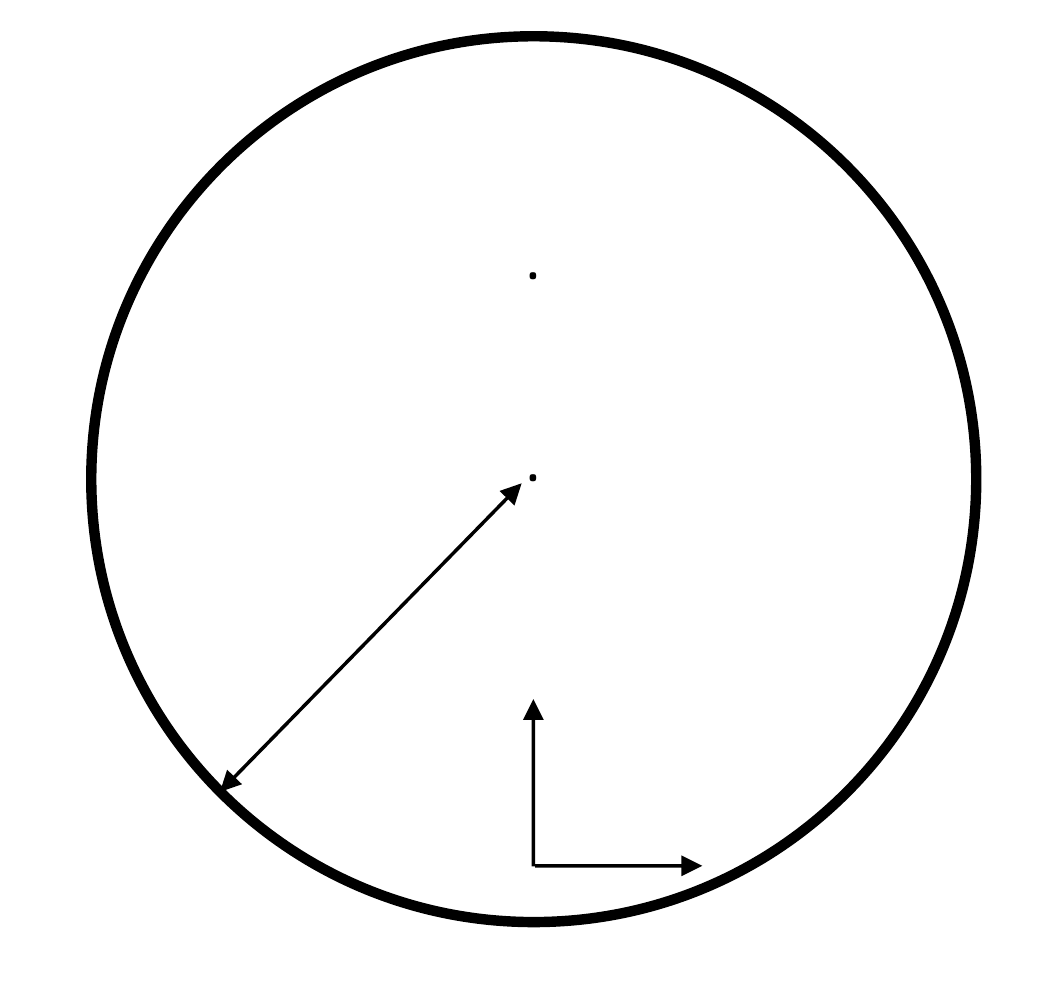}};

\node at (0.8,-1.55) {$x_c$};
\node at (0.26,-0.95) {$y_c$};
\node at (-0.85,-0.5) {$r$};
\node at (0.25,0.2) {$\textrm{T}_\textrm{x}$};
\node at (0.28,1.15) {$\textrm{R}_\textrm{x}$};

\end{tikzpicture}
\caption{The cross-section of a circular tunnel. Visible also are the position of the transmitter and the receiver.}
\label{fig:tunnel_configuration}
\vspace*{-0.55cm}
\end{figure}
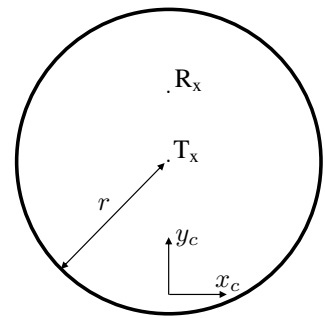

We will illustrate the key steps for formulating an ML propagation model, using an example of an ANN-based model predicting the RSS inside a straight circular tunnel. The ML model of our choice is the ANN. The cross-section of such a tunnel can be seen in Fig.~\ref{fig:tunnel_configuration}. The position of the transmitter $\text{T}_\text{x}$ remains fixed at the center of the tunnel, with coordinates $(x_c, y_c) = (0, r)$, where $r$ is the radius of the circular tunnel. The receiver $\text{R}_\text{x}$ is moving along the tunnel in the $z_c$ direction, with its position on the $x$-$y$ plane fixed at $(x_c, y_c) = (0, 3r/2)$. Thus, the receiver's trajectory is emulating an antenna, mounted on a moving train or wagon. The length of the tunnel is 500\,m. The transmitter radiates a 20\,dBm single Gaussian beam \cite{xhingqi_vpe}, while the receiver uses a half-wave dipole antenna. Both antenna gains are 1\,dBi.

Table~\ref{tab:inputs} lists all the input features used, as well as their respective range of values. Two different frequency bands are used, centered at 900\,MHz and 2.4\,GHz respectively, each with a bandwidth of 200\,MHz. Since these two frequency bands exhibit different propagation characteristics, separate ANNs are created for each band. Our particular selection of input features is based on our knowledge of parameters that influence the RSS. Both the geometry of the tunnel, the axial distance between the receiver and the transmitter, as well as the frequency of operation are such features. Thus, each input vector $\boldsymbol{x}$ is 3-dimensional.

\begin{table}[h!]
\vspace*{-0.2cm}
\centering
\caption{Circular tunnel input features}
\label{tab:inputs}
\scalebox{0.9}{
\begin{tabular}{ |c||c|c|c|c|c| } 
 \hline
 Parameter & Symbol & Min. Value &  Max. Value & Increment & \# conf.\\
 \hline
 Radius & $r$ & 2\,m & 4\,m & 0.1\,m & 21\\
 \hline
 1st Freq. Band & $f_1$ & 800\,MHz & 1\,GHz & 25\,MHz & 9\\
 \hline
 2nd Freq. Band & $f_2$ & 2.3\,GHz & 2.5\,GHz & 25\,MHz & 9\\
 \hline
 Axial Distance & $z_c$ & 0\,m & 500\,m & 1\,m & 501\\
 \hline
\end{tabular}}
\vspace*{-0.3cm}
\end{table}

The three datasets are generated as follows. The $i$-th input vector $\boldsymbol{x}_i$ is associated with a scalar target value $y_i$, corresponding to the received power (in dBm) at the specified point. The target values are generated using an in-house VPE solver that has been extensively validated \cite{xhingqi_vpe}. We generate 189 training/validation tunnel configurations for each of the two frequency bands. Each tunnel configuration corresponds to 501 different input vectors $\boldsymbol{x}_i$, one for each receiving point along $z_c$. From these 189 tunnel configurations, 20 are randomly chosen to form the validation set. Another 24, different than those comprising the training and the validation set, are randomly generated and used as test cases.

The input data are then pre-processed. In our case, they are normalized to have zero mean and unit standard deviation. To achieve this, the following transformation is applied to the $j$-th input feature:
\begin{equation}
 x_{norm}^{j} = \frac{x^{j}-\mu^{j}}{\sigma^{j}}
\label{eq:normalization}
\end{equation}
where $\mu^{j}$ and $\sigma^{j}$ is the $j$-th input feature's mean and standard deviation, respectively, computed only over the training samples. That helps the ANN speed-up training, as all the input features have comparable ranges. We choose a cost function that minimizes the squared error between the network's predictions and the target values and also apply $\text{L}_2$ regularization in order to penalize the complexity term of Eq.~(\ref{eq:gen_bound}). Thus, the in-sample mean squared error (MSE) is given by:
\begin{equation}
\mathcal{E}_{\text{in}}(\boldsymbol{w}) = \frac{1}{N} {||g(\boldsymbol{x})- \boldsymbol{y}||}_{2}^{2} + \frac{\lambda}{2}{||\boldsymbol{w}||}_{2}^{2}
\label{eq:cost_function}
\end{equation}
where $\lambda$ is a regularization parameter, $N$ is the number of input samples, also called the batch size, and $\boldsymbol{w}$ is a vector containing all the weights of the network. We use the identity linear function for the output node and tanh activations for the rest. Finally, we use the ADAM optimizer \cite{adam}, a more advanced version of the gradient descent algorithm, as our learning algorithm for minimizing the cost function.

\def\fw{6.1cm}
\begin{figure}
\vspace*{-0.3cm}
\centering
    {\includegraphics[width=\fw]{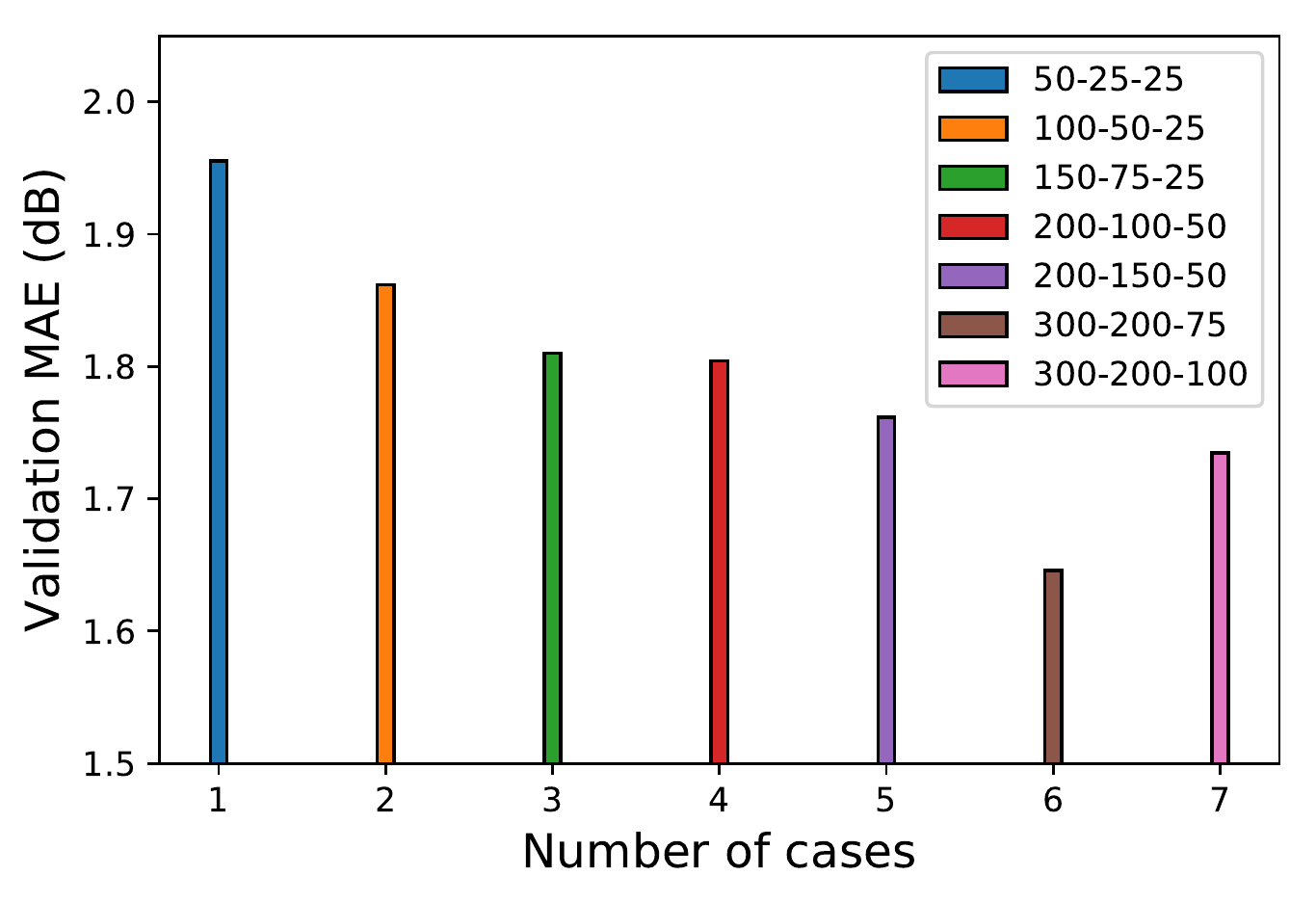}}
\caption{Hidden neuron selection in a 3-hidden-layer ANN.}
\label{fig:val_1}
\vspace*{-0.4cm}
\end{figure}

Before training the ANN, we choose the number of hidden layers and neurons per layer, evaluating different combinations of them with respect to their validation error. This is done by first training on the training set and then computing the MAE over all the validation samples for each different architecture. Then, we choose the architecture that gives the smallest MAE as our final model architecture. Fig.~\ref{fig:val_1} shows the validation MAE for a 3-hidden-layer ANN of various configurations of nodes per layer, for the 900\,MHz case. Fig.~\ref{fig:val_2} illustrates the same for a 4-hidden-layer ANN. Based on these figures, a 300-200-75 node configuration seems optimal for the 3-hidden-layer ANN, since the validation error increases thereafter. Meanwhile, the 4-hidden layer case exhibits a slightly smaller MAE for the node configurations checked. It also follows a decreasing trend as the total number of nodes is increased. As a result, we choose the 4-hidden-layer architecture. Using less than 3 hidden layers makes it more difficult for the network to capture some of the oscillations of the signal. On the other hand, using more than 4 or further increasing the number of neurons per layer, increases training time without substantial improvement in accuracy. The same procedure is also followed to determine all other hyperparameters, such as the amount of regularization and the batch size of Eq.~(\ref{eq:cost_function}).

\def\fw{6.1cm}
\begin{figure}
\centering
    {\includegraphics[width=\fw]{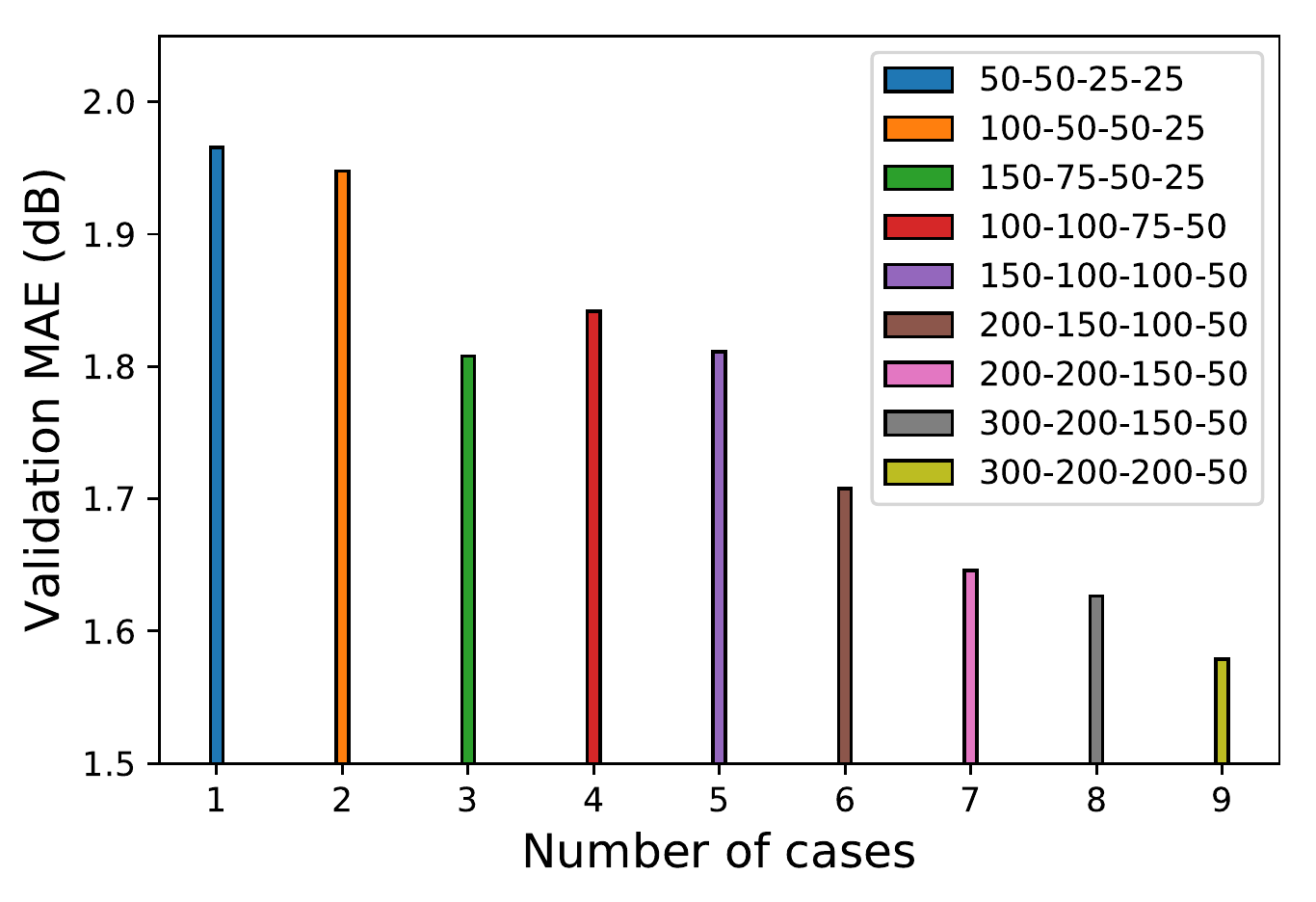}}
\caption{Hidden neuron selection in a 4-hidden-layer ANN.}
\label{fig:val_2}
\vspace*{-0.6cm}
\end{figure}

We then train the two ANNs. We set the same number of total iterations for both, and record the values of $\mathcal{E}_{\text{in}}$ on the training set. After a sufficiently large number of training iterations, the training error starts to converge while the validation error remains almost constant, allowing us to stop training. Then, we evaluate the trained model on the test set. The final training and test errors recorded for the two ANNs can be seen in Table~\ref{tab:errors}. It is noted that the 4-hidden-layer network is more accurate than its 3-hidden-layer counterpart. That is the reason we only trained the 4-hidden-layer ANN for the 2.4\,GHz band. Using the same set of hyperparameters and training time, we can see that all three errors increase, compared to the 900\,MHz case. That can be easily explained if we look closely at Figs.~\ref{fig:0.9_plot} and \ref{fig:2.4_plot}. The plots show the predicted RSS versus the actual one for two arbitrarily selected tunnels at the two different frequency bands. It is apparent that the ANN is very accurate at 2.4\,GHz too. The difference in the test MAE is mainly caused by the fast fading characteristics of the received signal. That is also validated by the large increase in the training MSE compared to the 900\,MHz case, since this type of error is much more sensitive to the oscillations of the received signal than MAE.

\def\fw{6.1cm}
\begin{figure}
\vspace*{-0.3cm}
\centering
    {\includegraphics[width=\fw]{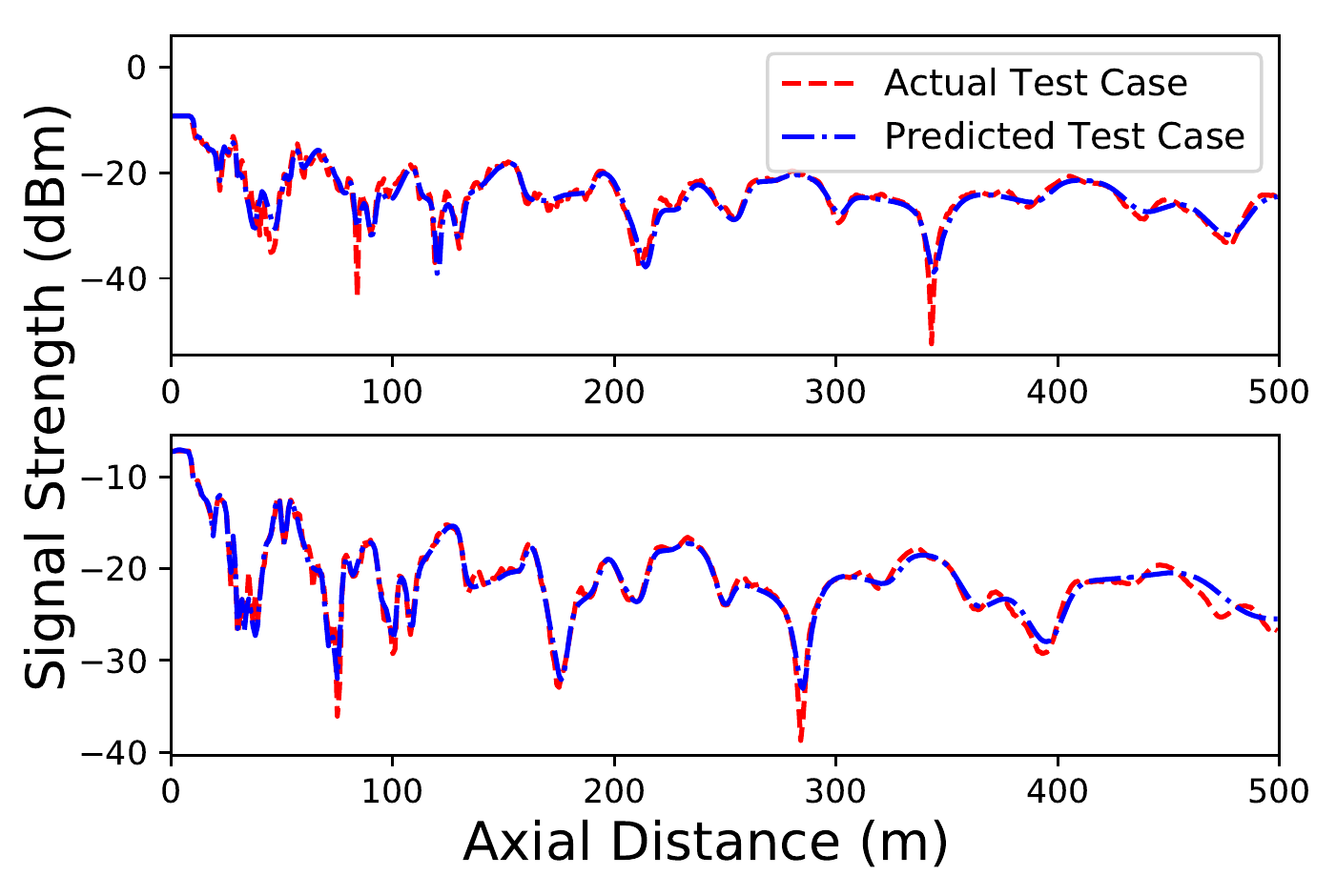}}
\caption{Predicted versus actual received signal strength of two arbitrarily selected tunnel configurations for the 900\,MHz case.}
\label{fig:0.9_plot}
\vspace*{-0.3cm}
\end{figure}

\def\fw{6.1cm}
\begin{figure}
\vspace*{-0.25cm}
\centering
    {\includegraphics[width=\fw]{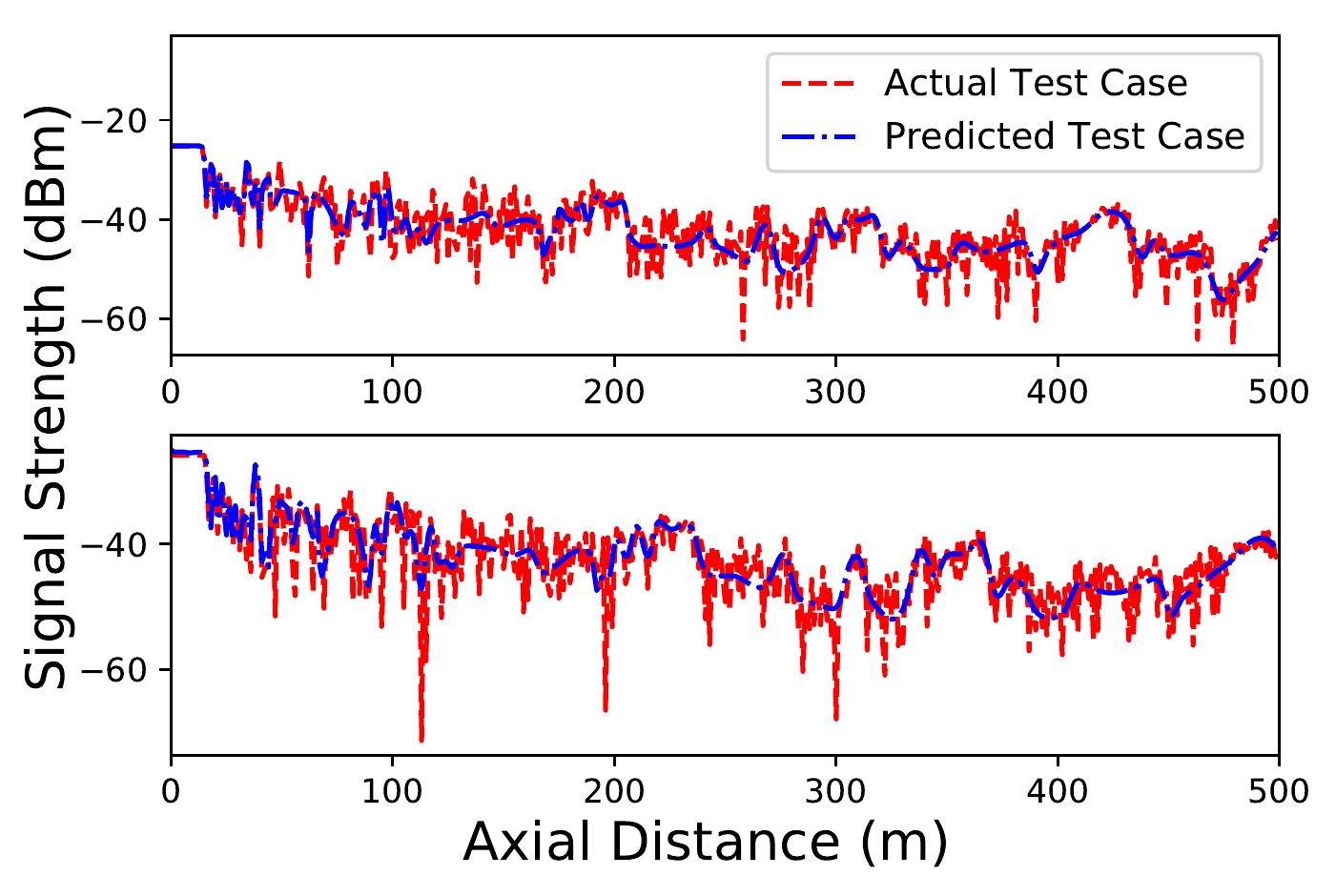}}
\caption{Predicted versus actual received signal of the same two arbitrarily selected tunnel configurations for the 2.4\,GHz case.}
\label{fig:2.4_plot}
\vspace*{-0.55cm}
\end{figure}

\begin{table}[h]
\vspace*{-0.2cm}
\centering
\caption{Error metrics for the case study}
\label{tab:errors}
\scalebox{0.92}{
\begin{tabular}{ |c||c|c|c|c| } 
 \hline
 Frequency & Layers & Train. MSE (dB) & Val. MAE (dB) &  Test MAE (dB)\\
 \hline
 900\,MHz & 3 & 2.677 & 0.946 & 0.827 \\
 \hline
 900\,MHz & 4 & 1.677 & 0.785 & 0.725 \\
 \hline
 2.4\,GHz & 4 & 15.03 & 2.579 & 2.746 \\
  \hline
\end{tabular}}
\end{table}

\vspace{-0.3cm}

\section*{Acknowledgements}
This work has been supported by the Huawei Innovation Research
Program (HIRP) and the Natural Sciences and Engineering Research Council of Canada's (NSERC) Discovery Grant. \\

\vspace{-0.6cm}

\bibliography{IEEEabrv, main.bib}
\bibliographystyle{ieeetr}


    
\begin{table*}[h]
\vspace{-0.4cm}
  \centering
  \begin{threeparttable}
  \caption{Type of input, ML model and output of papers under review.}

\begin{tabular}{ |C{1.3cm}||C{1.4cm}|C{0.8cm}|C{1.2cm}|C{0.8cm} |C{1.45cm}|C{9cm}| } 
 \hline
Paper  & Type of env. (U, sR, R)\tnote{*}& ANN-based model & Non-ANN based model & Hybrid model  &  Type of training data (Ms, Sn)\tnote{**} & Output\,-\,Application \\
  \hline
  


\cite{neskovic, popescu_1, anitzine, monteiro} & U  & MLP &  -  & -  & Meas. & The first paper is an RSS microcell prediction model at 900\,MHz, while the second one corresponds to a PL prediction model at 1890\,MHz. The third one is a PL prediction model at 900 and 1800\,MHz, while the last paper predicts the electric perimittivity and conductivity of the ground. \\
\hline


\cite{balandier, perrault} & U & MLP & - & \checkmark  & Meas.  & RSS prediction model at 170\,MHz, RSS error correction and  acceleration of an RT solver, respectively. \\
\hline

\cite{gschwendtner, cheerla, isabona} & U & MLP & - & \checkmark & Meas. &  PL correction of CWI (first two papers) and the log-distance model (third paper). The second paper operates at  800 and 1800\,MHz Comparison with empirical models. \\ 
\hline

\cite{ferreira, panda, ehbota} & U  & MLP &  -  & \checkmark  & Meas. & The first paper corresponds to a PL prediction model driven by the CKE empirical model at 1140\,MHz. The second paper implements PL correction of Hata model. The third one is a PL prediction model driven by an  Adaline network at 1900\,MHz. \\
\hline


\cite{sotiroudis_1, sotiroudis_3, sotiroudis_2, cerri} 
& U & MLP & - & - & Synth. (RT) & PL prediction models. In the last paper, the PL equation is divided into free space and building attenuation. The latter is computed by the ML model. \\
\hline


\cite{piacentini} & U & MLP & SVR & - & Meas. & PL prediction model. Experiments with PCA/nPCA. \\
\hline

\cite{moreta} & U & MLP & Various & - & Meas. & RSS prediction model for DTT systems. Comparison between MLP, $k$-NN, SVR, RF, AdaBoost, LASSO and kriging.  \\
\hline

\cite{lee, kuno} & U & CNN & - & - & Synth. (RT) & PLE prediction models; the first one at 28\,GHz. \\
\hline

\cite{ahmadien} & U, sR & CNN & - & - & Synth. (RT) & PL distribution estimation at 900\,MHz and 3.5\,GHz. The environments were constructed in RT from satellite images.\\
\hline

\cite{jang} & I & CNN & - & - & Meas. & User localization in an indoor environment by learning RSS maps.\\
\hline

\cite{imai} & U & MLP, CNN & - & \checkmark & Synth. (RT) &   Hybrid PL prediction model consisting of an CNN and an MLP-ANN. \\ 
\hline

\cite{timoteo} & U & - &  SVR  & -  & Meas. &  PL prediction model at 853\,MHz. Comparison with empirical models. \\
\hline

\cite{fernandes} & U & - & GA & - & Meas. & PL prediction model. \\
\hline
 
\cite{zhang_y} & U & MLP & SVR, RF & - & Meas. & PL prediction model at 2021\,MHz. \\ 
\hline

 
\cite{jo} & sR & MLP & - & - & Meas. & 
PL and shadowing prediction model at 450, 1450 and 2300\,MHz. \\ 
\hline


\cite{ostlin, lina_wu} & R & MLP &  -  & - & Meas. & The first two papers correspond to macrocell PL prediction models at 881\,MHz. They include comparisons with empirical models. The third paper is an RSS prediction model, using also PCA. \\
\hline


\cite{oroza} & R  & MLP &  RF, $k$-NN, AdaBoost & - & Meas. & PL prediction model for sensor network connectivity at 2.4\,GHz. Comparison between different ML models.  \\
\hline

\cite{ribero} & R & MLP, CNN & - & - & Meas. & PG prediction model at 1.8\,GHz  over irregular terrain. \\
\hline

\cite{cheng1} & R & WNN &  -  & -  & Synth. (VPE)  &  RSS prediction model for various frequencies (from 900\,MHz up to 2.2\,GHz). \\
\hline


\cite{thrane} & U, R & MLP, CNN & - & \checkmark & Meas. & PL prediction model at 811 and 2630\,MHz,  trained by measurements and satellite images. \\
\hline

\cite{zhang_j} & U, R & - & Various & - & Both (5G simulations) & 
Environment classification for outdoor railway environments.  Comparison between $k$-NN, SVR, $k$-means and GMM models. \\ 
\hline


\cite{popescu_2} & U, sR & RBF &  -  & \checkmark & Meas. & PL error correction of CWI model.  Comparison with empirical models.  \\
\hline

\cite{karra} & U, sR & - & Various & - & Meas. & PL prediction model trained by UAV-taken images.  Comparison between $k$-NN, SVR, RF, AdaBoost, GTB and VR.  \\ 
\hline


\cite{ayadi} & All & MLP &  -  & -  & Meas. &
Multiband (450-2600\,MHz), multi-environment PL prediction model. Comparison with empirical models. \\
\hline

\cite{gideon} & All & ESN & SVR & - & Meas. & 
RSS  prediction model at 900, 1800 and  2100\,MHz at various terrain types.  \\ 
\hline


\cite{zaarour} & - & MLP, RBF &  -  & -  & Meas. & Wideband PL prediction model at 60\,GHz in mines. \\
\hline

\cite{di_wu} & - & MLP & - & - & Meas. &  PL prediction model for railway  environments at 930\,MHz. \\ 
\hline

\cite{aris} & - & MLP & - & - & Synth. (VPE) & PG prediction model for tunnel environments at 0.9 and 2.4\,GHz. \\
\hline

\cite{wen} & - & MLP & SVR, RF, AdaBoost & - & Meas. & PL prediction model for aircraft cabin  at 2.4, 3.52 and 5.8\,GHz. \\  
\hline

\cite{adeogun, falcone} & I & MLP & - & - &  Synth. (PGr, SV / RT) & Statistical propagation parameters prediction for indoor model calibration at 60\,GHz (first paper). RT acceleration ML model at 2.4\,GHz (second paper). \\
\hline

\cite{hou} & I  & - & RF  & - & Synth. (RT)  &  RSS prediction acceleration of RT  for indoor environments. \\
\hline

\cite{adege, hoang, yadav} & U, I  & RNN & -  & - & Meas. &  The first paper is about user trajectory prediction at 860\,MHz based on measured RSS in urban environments. The other two, concern fingerprinting localization in indoor environments. The third paper used an LSTM RNN. \\
\hline

\cite{turabieh, hsieh} & I  & RNN & -  & \checkmark & Meas. & User localization in indoor environments. The first paper used two RNNs for classifying building and floor respectively, while the second paper two  LSTMs for calculating the coordinates and floor, respectively, of sensors. \\
\hline

\cite{xu, cheng2} & I / -  & RNN & -  & - & Synth. &  The first paper is about tracking in indoor environment. Their LSTM is trained by simulations. In the second one, an Elman RNN is used to calculate the propagation factor over flat earch.  \\
\hline

\cite{wang} & I  & RNN, CNN & -  & \checkmark & Meas. &  User location estimation in an indoor environment by a hybrid model (LSTM and ResNet). \\
\hline

\cite{chen, li} & I  & GAN, CNN & -  & \checkmark & Both &  CSI indoor localization. The first paper used a hybrid of a DCGAN sharing weights with a CNN classifier, while the second one a DCGAN with an MLP.  \\
\hline

\cite{belmonte} & I  & GAN, RNN & -  & \checkmark & Both &  User tracking in an indoor environment A cGAN was used for data augmentation, while an LSTM computes the location of the user. \\
\hline

\cite{gan} & I  & DBN & -  & \checkmark & Both &  User tracking in an indoor environment.\\
\hline

\end{tabular}
\begin{tablenotes}\footnotesize
\item[*] U: urban, sR: semi-rural, R: rural, I: indoor [**] Meas.: measured, Synth.: synthetic
\end{tablenotes}
\label{tab:1}
  \end{threeparttable}
\end{table*}


\end{document}